# A Systematic Literature Review of Test Breakage Prevention and Repair Techniques

Javaria Imtiaz, Salman Sherin, Muhammad Uzair khan, Muhammad Zohaib Iqbal
Software Quality Engineering and Testing (QUEST) Laboratory,
National University of Computer and Emerging Sciences, Islamabad, Pakistan
{javaria.imtiaz, salman.sherin, uzair.khan, zohaib.iqbal}@questlab.pk

**Abstract:**
**Context:** When an application evolves, some of the developed test cases break. Discarding broken test cases causes a significant waste of effort and leads to test suites that are less effective and have lower coverage. Test repair approaches evolve test suites along with applications by repairing the broken test cases.

**Objective:** Numerous studies are published on test repair approaches every year. It is important to summarise and consolidate the existing knowledge in the area to provide directions to researchers and practitioners. This research work provides a systematic literature review in the area of test case repair and breakage prevention, aiming to guide researchers and practitioners in the field of software testing.

**Method:** We followed the standard protocol for conducting a systematic literature review. First, research goals were defined using the Goal Question Metric (GQM). Then we formulate research questions corresponding to each goal. Finally, metrics are extracted from the included papers. Based on the defined selection criteria a final set of 41 primary studies are included for analysis.

**Results:** The selection process resulted in 5 journal papers, and 36 conference papers. We present a taxonomy that lists the causes of test case breakages extracted from the literature. We found that only four proposed test repair tools are publicly available. Most studies evaluated their approaches on open-source case studies.

**Conclusion:** There is significant room for future research on test repair techniques. Despite the positive trend of evaluating approaches on large scale open source studies, there is a clear lack of results from studies done in a real industrial context. Few tools are publicly available which lowers the potential of adaption by industry practitioners.

# 1. Introduction

Testing is an important activity to assure the quality of software systems. Modern software development practices like DevOps, continuous integration and development encourage automated execution of test cases by requiring test engineers to develop test scripts. This leads to several advantages such as automated test execution, test effort reduction, efficient usage of resources and an easy to use regression test suite. Availability of test scripts that can be executed automatically improves the efficiency of testing, but can lead to significant challenges in the maintenance of the test scripts. As the system under test (SUT) evolves, a number of test scripts can fail due to changes in the SUT. Therefore, the test suite needs to evolve along with the SUT.

For regression testing to be effective, the test suite must be updated to keep pace with the changes in SUT. Each introduced change may lead to failed test cases. Some tests fail due to the presence of faults or bugs in the application code. However, some tests may stop prematurely due to modifications in the application code such as repositioning or renaming of existing elements, locator and layout changes, etc. Such premature stopping of test cases due to changes in the SUT is referred to as *test breakage.* The broken test cases cannot be executed on the updated SUT without fixing the implementation of the test cases (the test scripts). The existing literature classifies the regression test suite as *usable, unusable* and *obsolete* test scripts [1]. The usable test cases conform to the existing functionality because they are not affected by the changes made in the evolved (modified) version of SUT. The unusable/broken test cases contain at least one statement that cannot be executed successfully. Such un-executable statement(s) may break the whole test case but the test case can be 'fixed' by applying repairing transformations to the test case implementation. Obsolete test cases fail to execute on the updated version and are not repairable, for example, they correspond to functionality that has been removed from the SUT. The changes that break test cases can be *structural* or *logical* [2]. Structural changes deal with the layout and structure of the application whereas the logical changes deal with modification in business logic or functionality.

Discarding broken test cases after modifications highly affect the quality of the regression test suite. This reduces the size of the test suite and requires significant effort in re-writing and re-recording test scripts from scratch. Even small modifications can lead to a large number of broken test cases, in some cases up to 74% of the test suite [3]. Discarding the broken test cases therefore leads to a significant increase in the cost of testing and may reduce the quality of the test suite. Therefore, repairing broken test scripts is an important task [4].

Over the past decade, researchers and practitioners have proposed different techniques for automated repairing of broken test scripts of evolving software systems [3, 5-7]. Broadly, the test script repair approaches perform three essential steps, (i) examine and classify difference between the original and modified versions of the evolving SUT, (ii) identify the broken test scripts, (iii) repair the broken test scripts using repairing transformations.

In this paper, we systematically identify, summarize and evaluate the existing literature to find gaps in the area and to position new research activities. We present a Systematic Literature Review (SLR) [8] in which we review 41 papers on test breakage prevention and automated repairing of test scripts. SLRs are used to investigate, categorize, and evaluate the existing literature in a particular research area by applying well-defined inclusion and exclusion techniques. The contribution of this study is twofold. First, it helps new researchers in a structured understanding of the area by indexing the existing studies and by providing new research directions. Second, it helps practitioners to understand state-of-the-art tools, techniques and their appropriate usage. More specifically, we provide the following contributions in the area of test script repair:

- We identify the test case repair approaches presented in the literature and classify the studies in terms of the type of contribution made, kind of approaches, and testing frameworks used for writing and recording test scripts.
- We classify the causes behind test case breakages that are presented in the literature and provide a taxonomy of commonly identified changes that can result in test breakage.
- We report on the evaluation of test breakage prevention and test script repair techniques. We document any identified empirical evaluations, benchmark case studies and the widely used metrics to evaluate the quality of the proposed techniques.
- We provide implications of existing test repair techniques for practitioners based on the available evidence on the application of test repair techniques and tools.
- Finally, we offer new directions for future research by identifying the gaps in the area.

The rest of the paper is structured as follows. Section 2 presents the research methodology and the research questions used in this study. Section 3 discusses the answers to our research questions and the results of the review. Section 4 presents a discussion on our findings and take away for researchers and practitioners. Section 5 presents the related work and section 6 discusses different threats to validity. Finally, Section 7 concludes the paper with a discussion of potential future directions.

## 2. Research Methodology

We perform a Systematic Literature Review (SLR) by following widely accepted guidelines given in [8-10]. Based on the guidelines, given in [9], we conducted this study in three steps, i.e., Planning, Conducting and Reporting. To clearly articulate the aims of the study, we use the Goal-Question-Metric (GQM) paradigm, given in [10]. Table 1 depicts our review protocol for conducting this SLR.

Table 1 Research protocol used in this study

| Phases | Steps |
|---|---|
| Planning | • Goals<br>• Research Question<br>• Selection of Online-Digital Libraries<br>• Formulation of the Query String<br>• Definition of Inclusion and Exclusion criteria |
| Conducting | • Study Selection<br>• Metrics/Attributes identification<br>• Data Extraction<br>• Data Synthesis |
| Reporting | • Dissemination of results<br>• Report formatting |

## 2.1 Planning the Review

### 2.1.1 Goals

The aim of this review is to identify, review and synthesize the current state-of-the-art in the field of test case evolution. We aim to identify the recent trends and limitations, to evaluate the maturity of the area and discuss the opportunities for future research from the point of view of researchers and practitioners. Based on the objective of the study, we identify the following research goals:

**G1:** To systematically map (classify) the state-of-the-art in the area of test case breakage prevention and test case repair.
**G2:** To study the common changes or causes of test case breakages in evolving applications.
**G3:** To study the nature of the published evidence on the effectiveness of the approaches, their evaluation, the tools being used, and subject applications.

Goals **G1**, **G2**, and **G3** focus on gathering in-depth knowledge of test case repair research and empirical evaluation(s) performed to validate the proposed approaches. Based on the aforementioned goals of the study, we have formulated and grouped our research questions in five categories. Research questions in each category are further decomposed into multiple sub-research questions to rigorously extract and analyze the information.

### 2.1.2 Research Questions

**RQ 1:** What is the current state-of-the-art in the field of test case repair? The RQ is further divided into sub-questions as follows:

- **RQ 1.1: Type of research contribution*:*** What are the contributions of different studies in the field of preventing test case breakage and test case repair and how many studies present techniques, tools, frameworks, guidelines, and processes? To answer this question, we have adopted the classification proposed by Petersen et al. in [12] by extracting contribution facet from each paper and classifying the paper in the corresponding class.
- **RQ 1.2: Type of research method:** What type of research methods have been used in the published studies on test repair? We answer this aspect of research by using the guideline of Petersen et al. [12] to classify the research approach of studies. Each paper is placed in one or more of the following categories: validation research, evaluation research, solution research, opinion research and experience research.
- **RQ 1.3: Test case repair tools:** What tools exist to repair the broken test cases of evolving applications? Availability of tools has important implications for practitioners. The answer to this RQ provides a list of test case repair tools developed and used in the studies.

- **RQ 1.4: Test frameworks:** How many of the techniques are specific to certain testing frameworks and how many are test repair techniques at a generic level that can be applied to any testing framework? This RQ classifies the techniques as generic solutions that are not tied to a particular framework and others which are tightly coupled with certain testing frameworks such as Selenium, JUnit, etc.
- **RQ 1.5: Automation level:** What is the automation level of techniques proposed in the area? It classifies whether proposed techniques are manual, automatic or semi-automatic (requiring some manual intervention).
- **RQ 1.6: Type of approaches used to deal with test repairs:** What type of approaches have been used to deal with test case repair? These approaches can be classified into broader categories such as model-based approach, search-based approach, and heuristics-based approaches, etc. However, we allow for overlap between the categories.

**RQ2: Causes of test case breakages:** What are the common causes of test case breakages in the evolving applications? In this RQ, we investigate the common causes of test case breakage identified in the included literature. The answer to RQ provides a taxonomy of causes of broken test cases that are reported in the literature.

**RQ3:** What types of SUT have been used for the evaluation of test case repair and breakage prevention approaches? In answer to this research question, we list the case studies used for the evaluation of techniques proposed in the covered primary studies and to identify if they are academic, open source or industrial case studies. To answer this question, we have formulated the following sub-questions:

- **RQ 3.1: Characteristic of SUT:** What is the type, scale and size (in terms of LOC) of each software system whose test cases are being analyzed in the included studies?
- **RQ 3.2: Type of metrics:** What are the metrics used for assessing the cost-effectiveness of test case repair approaches?
- **RQ 3.3: Share of industrial case studies:** What percentage of work cite evidence from applying the approaches on real industrial case studies? We differentiate between evidence from open source case studies (which might also be used commercially but are analyzed in lab settings) and evidence from evaluations in an actual industrial context.

### 2.1.3 Selection of online-digital libraries

A search for the relevant articles was carried out to answer the research questions. We focused on major digital libraries (given in Table 2) and augmented the search process using a well-defined methodology of snowballing used by other studies [13-16]. For snowballing we follow the guidelines by Wohlin et al. [17]. Suitable repositories were identified based on previous research experience and suggestions provided by Chen et al. [18]. The automated search process resulted in a number of duplicate studies in the initial search but we preferred a conservative approach over reducing redundancy in the initial search results. The search query formulation is discussed in subsection 2.1.4.

Table 2 Digital libraries and search engines

| Source | URL |
|---|---|
| Google Scholar | https://scholar.google.com.pk/ |
| IEEE Xplore | http://ieeexplore.ieee.org/ |
| ACM Digital Library | http://dl.acm.org/ |
| Springer Link | http://link.springer.com/ |
| Wiley Online Library | http://onlinelibrary.wiley.com/ |
| Science Direct | http://www.sciencedirect.com/ |

### 2.1.4 Formulation of query string

In order to include relevant publications in the pool of papers, all authors of this paper identified and proposed potential search keywords in several iterations. We performed keyword-based article extraction which provides relevant results. The search string was formulated through the following steps:

1. Identify search keywords from research questions.
2. Identify search keywords in relevant paper's titles, abstracts.
3. Identify synonyms and alternative words of search terms.
4. Connect identified keywords using logical ANDs and ORs operators.

Following keywords and their synonyms are identified (after consolidating the suggestions of all authors) to formulate the query string: (test case, test suite, test scripts, repair, co-evolve, maintenance, broken, unusable, obsolete). All synonyms were linked by inserting OR operator and different search terms were connected through AND operator. The final main query string is as follows:

***("Test case" OR "test suite" OR "test script") AND ("repair" OR "coevolve" OR "maintenance") AND ("broken" OR "obsolete" OR "unusable")***

The different variations of formulated search string were then provided to six search engines for an automated search. Search was performed on full text according to the guidelines provided by each database. Additionally, we also perform manual searching to mitigate the risk of missing articles. The manual search includes the following steps.

- We verified the selection of studies by cross-checking the references of the papers.
- The personal web pages and Google Scholar profiles (where available) of active researchers of the area were visited.
- The publication archives of specific venues where the higher number of papers is published (from the initial set of retrieved primary studies) were explored.

### 2.1.5 Defining inclusion and exclusion criteria

To select the relevant papers, we developed inclusion and exclusion criteria. We applied the criteria to the studies retrieved in the previous phase of source selection by reading the title, abstract and keywords of the studies. Each paper was reviewed by at least two authors of this paper before inclusion or exclusion into the final selection. Any conflicts in the inclusion and exclusion of studies were resolved through multiple group discussions and review meetings. After the application of inclusion and exclusion criteria, 41 studies were retained for analysis out of 589 total studies. The details of study selection are given in section 2.2.1.

Our inclusion and exclusion criteria are as follows:

- IC1: Studies which propose any technique, framework or tool for test case repair and breakage prevention.
- IC2: Studies written in English.
- IC3: Studies which are available in full texts.
- IC4: Studies which are available in multiple versions, only the most recent was included.
- IC5: Studies which are peer reviewed.

The exclusion criteria are:

- EC1: Studies that do not propose any technique, framework or tool for test case repair and breakage prevention.
- EC2: Studies not written in English.
- EC3: Studies not available in full text.
- EC4: Duplicate studies were removed.
- EC5: All presentations, magazine articles, tutorials, lecture notes, editorials and other non-peer reviewed articles.

## 2.2 Conducting the Review

### 2.2.1 Selection of studies

Initially, we retrieved a total of 589 studies from the digital search by applying the query strings. At first step, duplicate studies (i.e., a paper present in more than one database) were removed from the initial pool of studies (IC4, EC4), resulting in the removal of 176 duplicate papers. In the next step, irrelevant literature was removed from the remaining set of 413 studies on the basis of title and abstract reading (IC1, EC1, IC2, EC2), which resulted in remaining 213 studies. Consequently, we have removed grey literature (presentations, magazine articles, tutorials, lecture notes, editorials and other non-peer reviewed articles) and studies not available in full text by thoroughly reading the introduction and conclusion of the papers (IC3, EC3, IC5, EC5) resulting in a total of 39 studies. To reduce the bias in the selection of studies, the first two authors performed a selection of studies independently and the results were then matched. Any disagreements between the authors in the selection of studies were discussed and resolved in follow up meetings by all authors where each author presented arguments for including or excluding a study. To further reduce the risk of missing any relevant work, the last two authors of this paper performed snowballing following the guidelines given by Wohlin et al. [17]. Snowballing resulted in two more papers being included in the final set of 41 studies for further analysis. Figure 1 illustrates the protocol of study selection.

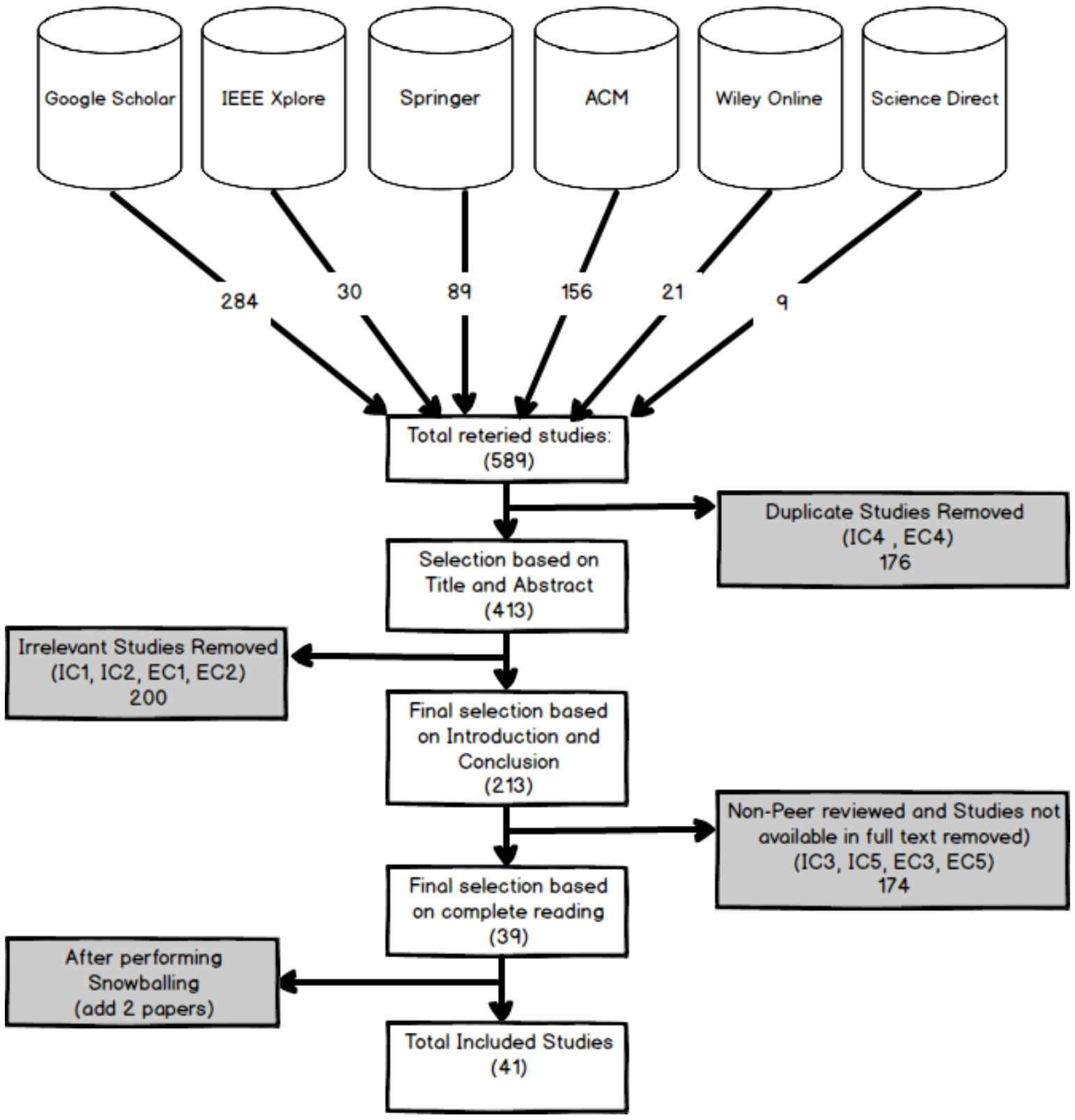

Figure 1: Protocol for study selection

### 2.2.2 Metrics

We performed a comprehensive analysis of the included studies to collect data to answer the research questions. Initially, we have defined the metric through research questions and then data was extracted against each metric from the papers and recorded in the spreadsheets. Subsequently, we maintained a data extraction form where we record the data against each attribute. Each study was reviewed at least by two reviewers (authors of the current study). Any conflicting papers were discussed with the third author (acting as tie-breaker) before the final decision was made. Table 3 shows the mapping of research questions with the identified metrics. Columns of the Table show goals, research questions, metrics and their possible values. We have identified the following metrics from the research questions such as ‘Type of research contribution, ‘Type of research methods, ‘Test case repair tools', ‘Test frameworks, ‘Automation level', ‘Type of approaches used to deal with test repairs’, ‘Causes of test case breakages’ and ‘Characteristic of SUT’.

### 2.2.3 Type of Research Contribution (corresponding to RQ 1.1)

Our initial goal was to identify the nature of articles in this domain and we extract the type of contribution made by each article. Possible values for this metric include technique, tool, framework, suggestions and processes [15]. This metric will help us to identify the distribution of effort between developing new test case repair techniques and tools.

### 2.2.4 Type of Research Method (corresponding to RQ 1.2)

This metric was used to access the type of research method used in each paper and is adapted from Peterson et al. [19]. It will help in identifying the maturity of the field that whether the papers have proposed some solutions without extensive validations or evaluated their approach through rigorous empirical methods. The possible values for this metric can be:

- **Solution Proposal:** A novel solution was proposed for a particular problem and its applicability was evaluated on a small case study.
- **Validation Research:** A novel technique was proposed and validated in a lab setting through an experiment.
- **Evaluation Research:** A novel technique was evaluated comprehensively through extensive experiments.

### 2.2.5 Test Case Repair Tools (corresponding to RQ 1.3)

We extracted the information regarding test case repair tools presented in each article. We assessed whether studies have proposed a new tool or extended existing tools which were developed in their previous works. We have also identified the developed tools and checked whether or not these tools are publicly available for download. This metric will present information about the test case repair tools developed and proposed in the area of test case repair and breakage prevention that are available to practitioners.

### 2.2.6 Test Framework (corresponding to RQ 1.4)

The metric 'Test framework' was used to extract data about which type of test cases are repaired by each technique. This metric will identify the most popular testing frameworks and tools in the area, for example, Selenium, QTP and JUnit. Furthermore, we have classified the frameworks on the basis of broad platforms, i.e., Mobile, Web and Desktop.

### 2.2.7 Automation Level (corresponding to RQ 1.5)

This metric is used to extract the automation level of each published technique. It will help to identify whether each proposed technique is manual, semi-automated or automated for repairing the broken test scripts.

- **Manual Approaches:** Test case repair approaches which are fully tester-assisted
  - Manual identification of a correspondence between the broken element and the test breakages.
  - Perform manual actions to fix the broken test scripts.
- **Semi-Automated Approaches:**
  - Automatically identify the correspondence between modified and new elements.
  - Perform manual actions to fix the broken test scripts.
- **Automated Approaches**:
  - Automatically detect the occurrence of breakages.
  - Automatically generate potential test fixes.
  - The validation of potential fixes may be manual.

### 2.2.8 Type of approaches used to deal with test repairs (corresponding to RQ 1.6)

This metric is used to assess the type of approach used in the papers. These approaches can be model-based, search-based, heuristic-based, computer vision-based, symbolic and concolic execution-based. This metric will identify the popular approaches used by different test case repair techniques. We allow for potential overlap between the categories.

- **Model-based Approach:** This category contains those studies which have used behavioral models to repair the test scripts. These models can be UML diagrams, control, and event flow graphs etc.
- **Search-based Approach:** This category contains studies which have used meta-heuristic algorithms (such as evolutionary algorithm e.g. genetic algorithm) to repair the broken test scripts.
- **Heuristic-based Approach:** This category contains approaches to problem-solving which is a practical method but not guaranteed to be optimal, still sufficient for the immediate goal of repairing test cases.
- **Computer Vision-based Approach:** This category contains studies which have used image recognition techniques to identify and control GUI components.
- **Symbolic and Concolic execution-based Approaches:** This category contains studies which have used some kind of program analysis techniques (such as static and dynamic code analysis) for repairing the test scripts.

### 2.2.9 Causes of test case breakages (corresponding to RQ 2)

Each study targets some specific set of changes for repair. We have extracted all identified changes from the studies which can cause test breakages. Each study aims to repair test scripts for a specific domain, such as web applications, mobile apps etc. We organize the causes of test breakages with respect to domains in Section 3.7. Each change is identified as structural or logical [2].

- **Structural Changes:** Structural changes affect the layout, and appearance of the application.
- **Logical Changes:** Logical changes affect the business logic of the application.

### 2.2.10 Characteristic of SUT (corresponding to RQ 3.1. RQ 3.2, RQ 3.3)

We collected the information related to the SUT (used for the evaluation or validation of approaches) in each of the included studies. Possible values for this metric are the number of subject applications, their names, size of SUT (in LOC), language and nature of SUT (such as open-source, industrial or a toy case study. We also identify the common metrics used for assessing the cost-effectiveness of test case repair approaches.

Table 3 Systematic map developed and used in our study

| Goals | Research Questions | Metrics | Possible outcomes |
|---|---|---|---|
| Goal 1 | RQ 1.1 | Type of research contribution | Technique<br>Tool<br>Taxonomy<br>Framework<br>Processes<br>Guidelines |
| | RQ 1.2 | Type of research method | Solution Proposal<br>Validation Research<br>Evaluation Research |
| | RQ 1.3 | Test Case repair tools | Type of Test Framework<br>Type of Repairs/Modifications<br>Test Case Execution<br>Language<br>Available for download<br>Published Year |
| | RQ 1.4 | Test framework | Selenium<br>JUnit<br>Selenium WebDriver<br>QTP |
| | RQ 1.5 | Automation Level | Manual<br>Semi-Automated<br>Automated |
| | RQ 1.6 | Approach Used | Model-based<br>Search-based<br>Heuristic-based<br>Computer Vision-based<br>Symbolic and Concolic Execution |
| Goal 2 | RQ 2 | Causes of Test case Breakages | Code level changes<br>Web GUI level changes<br>GUI level changes for desktop applications<br>Mobile GUI level changes |
| Goal 3 | RQ 3.1,<br>RQ 3.2,<br>RQ 3.3 | System Under Test | Name of SUT<br>Size of SUT (LoC)<br>Description of SUT<br>Frequency of the SUTs used in studies<br>Type of the SUT (i.e. open source, experimental)<br>Language<br>Application Domain (web, mobile, desktop)<br>Metrics |

# 3. Results and Discussion

In this section, we answer each of our research questions by using the extracted data.

## 3.1. Type of Research Contribution (RQ 1.1)

Overall, we identified 41 relevant studies from the selected sources, as shown in Table 13. Figure 2 shows the division of studies based on the type of contribution for all the 41 included studies in this paper. 39 studies proposed test case repair and breakage prevention techniques, 15 studies contributed test repair tools and two studies contributed frameworks, for example, S1 and S2. This shows that most of the research work is focused on contributing new techniques or improving previous techniques. Some of the papers were classified in more than one class, for example, S4 contributed a tool as well as a technique. The 15 studies (about 36%) which contributed tools also proposed techniques and therefore are classified under two classes, i.e., test case repair technique and test case repair tool. For example, S9 proposed a test repair technique and also developed a tool called ATOM. Section 3.3 provides a detail discussion on test case repair tools proposed in the included studies.

Figure 3 shows a different classification of the contribution made by the included studies. The existing literature can be categorized mainly into three classes, i.e., (i) studies that discuss mechanisms for avoiding test breakages, (ii) studies that discuss detection approaches for broken test cases, and (iii) studies that present approaches for test case repair. For example, S1 can be classified under ‘Test breakages repair' as it provides the technique for repairing broken test cases. S36 discusses test breakages detection and is classified under ‘Test breakages detection'. S28 discusses mechanisms for the avoidance of such breakages and therefore is classified under ‘Test breakages avoidance'. Some of the studies like S27 presented avoidance as well as a detection mechanism for test breakages and are classified in both the classes. Similarly, S5 presented avoidance as well as a repair mechanism for broken test cases, therefore, classified under repair and avoidance techniques. It can be seen that 29 (about 70%) studies discussed test repair mechanisms whereas nine studies (about 21%) discussed test breakages avoidance mechanisms and a few studies (5 out of 41, about 12%) examined detection of broken test cases.

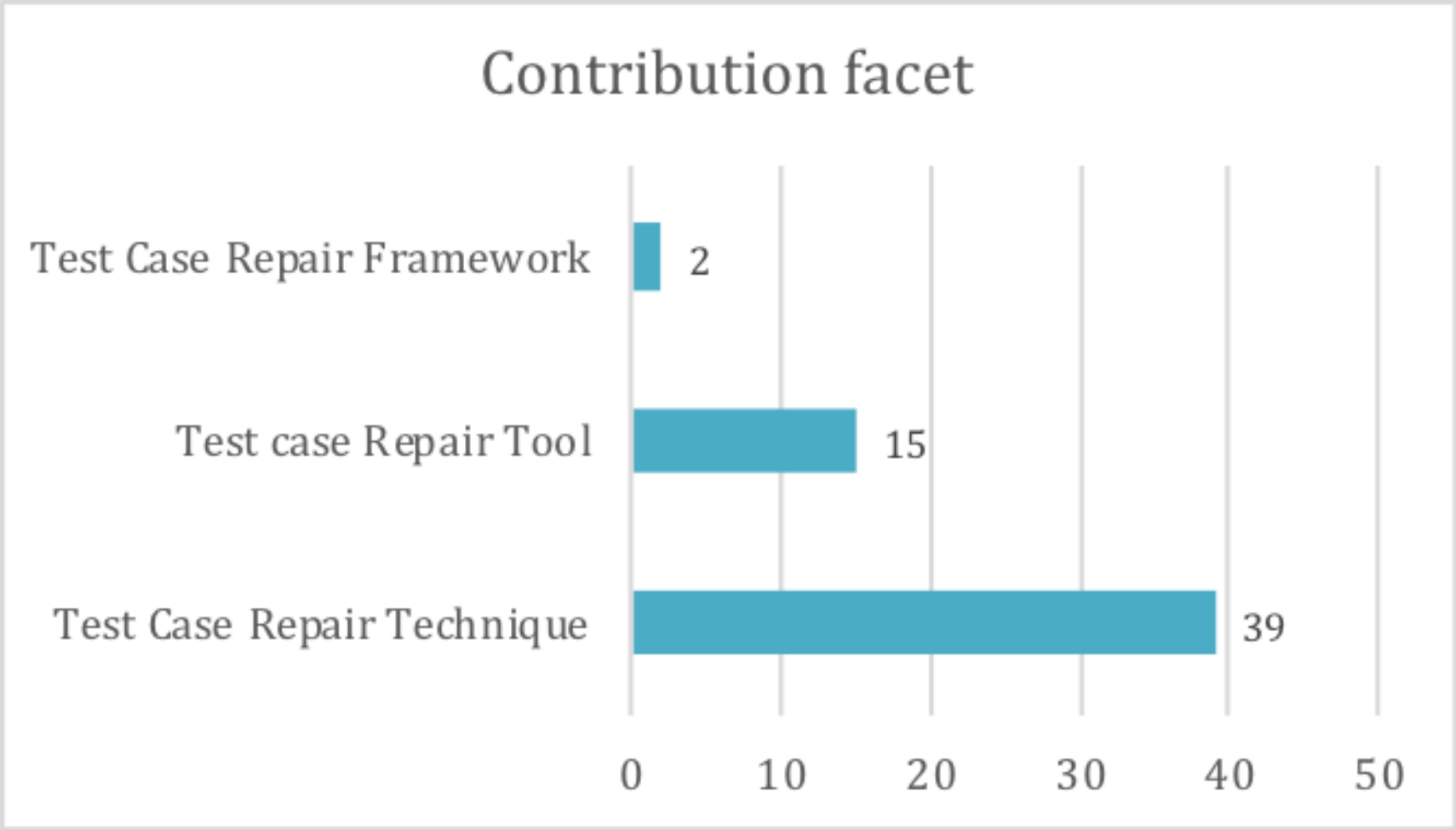

Figure 2 Type of contributions vs. number of papers

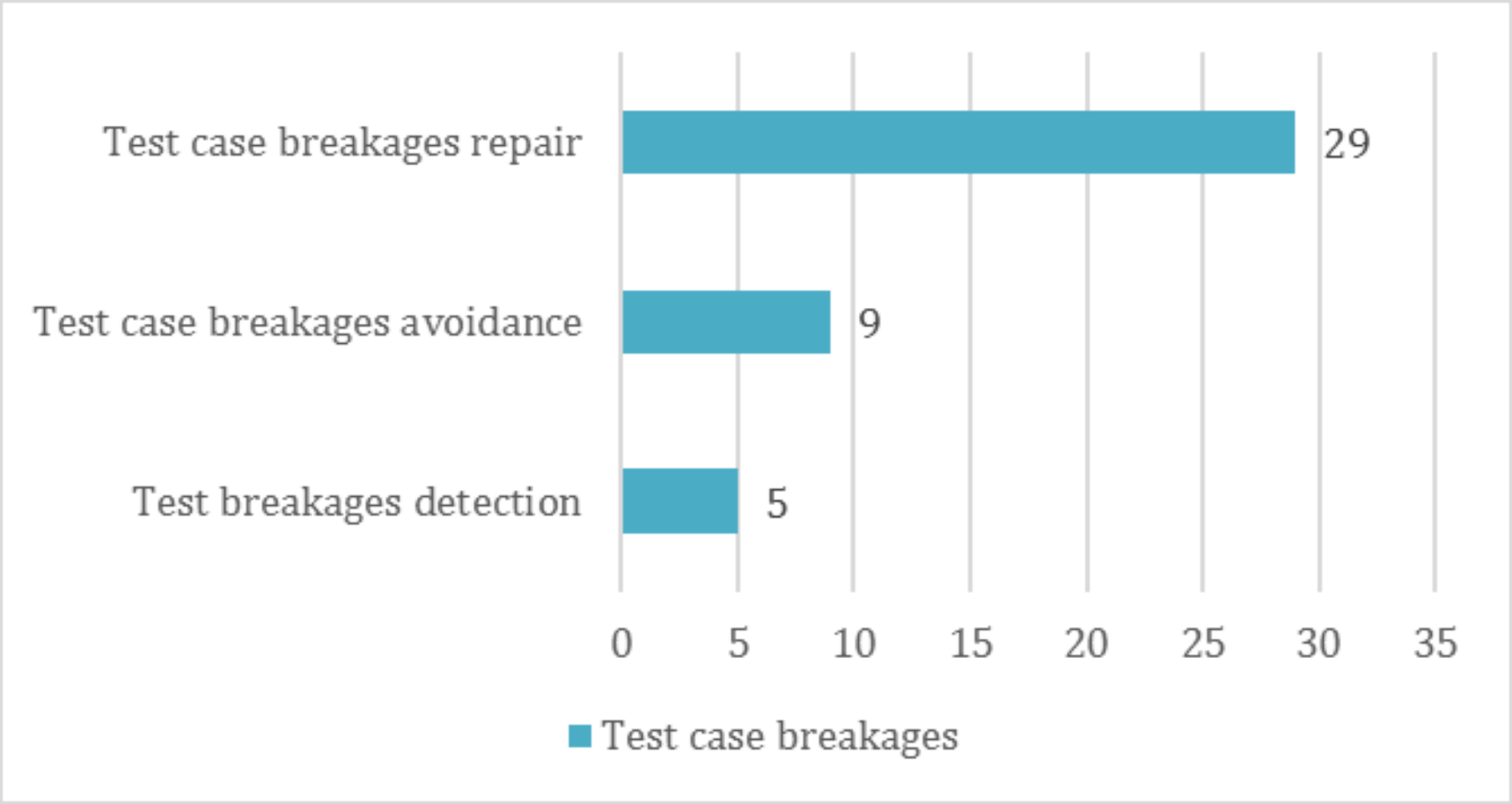

Figure 3 Classes of test case breakages vs. number of papers

## 3.2. Type of Research Facet (RQ 1.2)

Figure 4 shows the distribution of studies by research facet. In the area of test case repair, most of the research is dominated by validation research, about 63% (26) of the studies are mapped to validation research. This shows that studies are not only proposing the solutions but are also its applicability and effectiveness on subject applications. For example, S12 provides an automatic repair approach, implemented in a tool called TestCareAssistant and is evaluated by applying to the test cases of six different subject applications. There is a reasonable share of studies (6 studies, 14%) that are mapped to evaluation research. For example, S18 provide a GUI test script repair technique implemented in a tool called SITAR, which is extensively evaluated on open source subject applications by providing limitations and benefits of the proposed technique. Moreover, 21% (9) studies are categorized as solution research, for example, S3 provides an approach evaluated on small case studies. Validation research is more popular in the area which shows higher attention towards sufficient empirical evaluations conducted by the papers. We also found a number of works that focus entirely on empirical evaluations of test breakage prevention and test repair approaches. These are discussed separately in the related works Section.

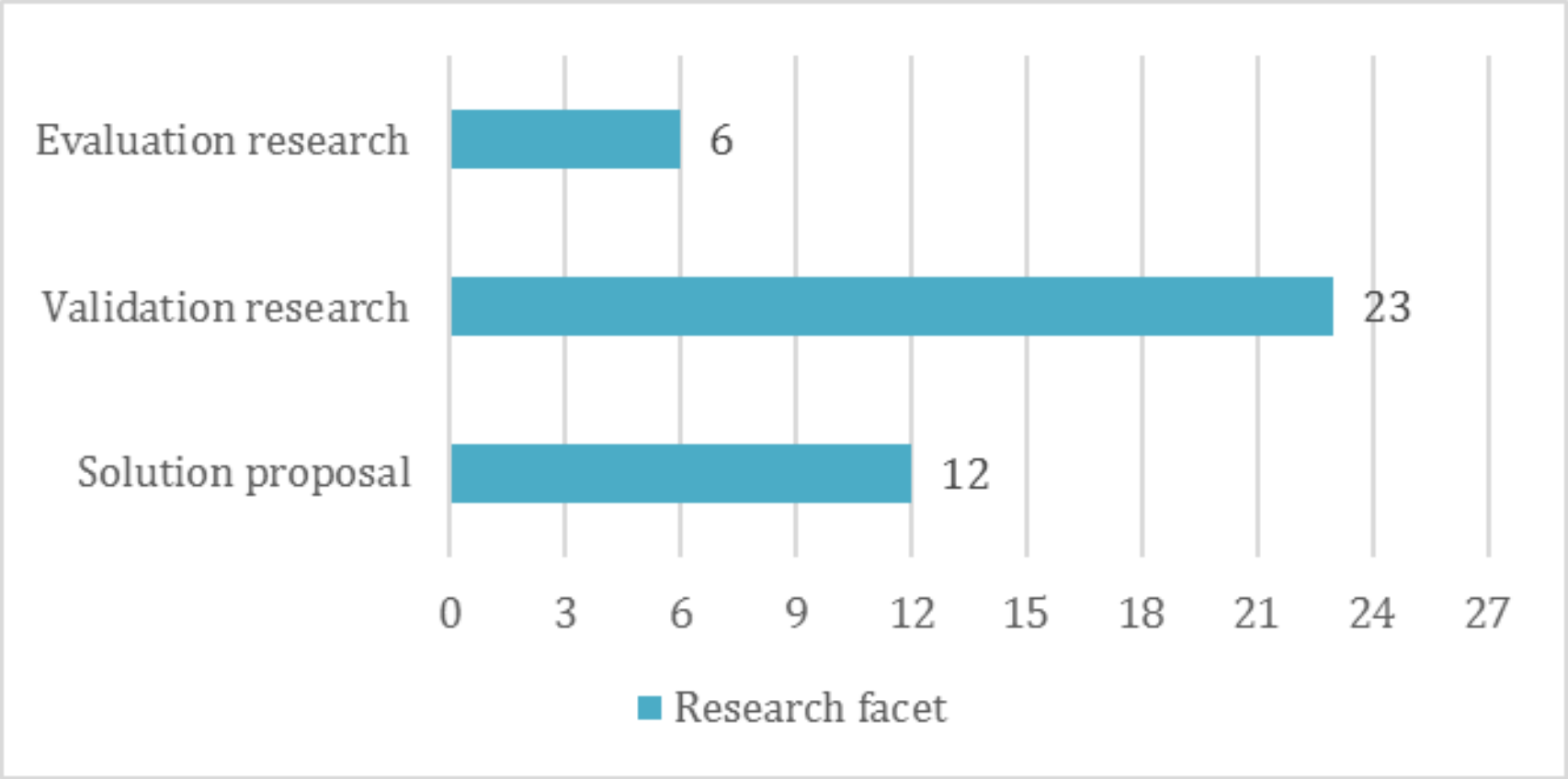

Figure 4 Research facet vs. number of papers

## 3.3. Test Case Repair Tools (RQ 1.3)

The development of automated tools is important for the transformation of academic research into its practical application in the industry. Without such automated tools test case repair techniques face significant challenges in industrial adaption. Table 4 summarizes different characteristics of existing test repair frameworks and tools developed for different types of applications. Overall, 15 tools are listed in which five tools (S2, S8, S10, S18 and S34) provide repairs for test cases of GUI-based application. Furthermore, five tools (S4, S5, S19, S23 and S39) provide repairs for code-based changes, two tools (S9 and S38) provide repairs for broken test scripts of mobile application and three tools (S20, S21 and S37) repair unusable test scripts for evolving web applications. We also searched these tools online, to check, whether they are available for the use of other researchers and practitioners. We conducted an online search on May 01, 2018 for tools where the authors explicitly mention that the tool is available for public download. Surprisingly, only four (S5, S19, S23, and S37) out of 15 tools were available for download.

Tools that are available for repairing the breakages of GUI test scripts are REST, GUIAnalyzer, Maintenance tool (called as maintenance tool by the study), FlowFixer and SITAR. S34 presented a tool called REST which is used as a plugin for eclipse to maintain and evolves GUI test scripts to test new versions. S2 proposed a Java-based tool, GUIAnalyzer, to provide a general solution for GUI test case maintenance by using the set of heuristics. S8 contributes a maintenance tool to automatically repair the GUI test scripts without any human intervention. Another study S10 provides GUI test evolution using FlowFixer, which suggests replacement actions for broken workflows. S18 presented a tool named SITAR uses a model-based technique to iteratively repair the obsolete low-level QTP scripts.

Tools such as ReAssert (S23), TESTEVOL (S5), TestCareAssistant (S19), TestFix (S4) and ITRACK (S39) are proposed to repair broken JUnit test scripts. Most of the tools from this category are focused on repairing the failing assertions. ReAssert suggests repairs in failed test scripts such as to replace literal values, change assertion methods, or replacing one assertion with several to pass the test. TestCareAssistant automatically repairs test cases broken by altering method signatures, by changing the number or type of the input parameters of the method. TestFix uses search-based algorithms to repair the broken JUnit tests by adding or deleting method calls. TESTEVOL enables the test-suite evolution and repairs the JUnit test cases by automatically applying test addition, deletion, and modifications without any human assistance. ITRACK matches the entities between two versions and identifies the existing test suite that needs to be changed to fix broken method calls by replacing the entities.

Tools that are available for repairing the breakages of mobile applications ATOM and CHATEM. S9 developed a tool ATOM to automatically maintain GUI test scripts of mobile apps for regression testing. S38 proposed a java-based tool, CHATEM, automatically extracts the changes between the two GUIs and generates maintenance actions for each change. Tools available to support web test breakages are WATER (S21), WATERFALL (S20) and VISTA (37). These tools are used to suggest potential repair actions for broken test scripts of capture-and-replay tools. WATER uses the browser's DOM tree to repair the broken Selenium test scripts for evolving web applications. It analyses the difference between two test executions, and then suggests repair for broken test scripts. WATERFALL uses WATER approach to repair the breakages due to the intermediate commits between the two major releases of web applications. VISTA repairs the DOM-based locators in web tests. It does so by tracking the broken web element across application versions using its visual appearance through the application of computer vision.

The growing trend of test case repair tools can be seen in the final column of Table 6. Most of the tools based on differential testing which executes whole test suite on both the original and modified version of applications for identification of broken or failed test cases. Such techniques have a higher execution cost as they require all test cases to be executed for the identification of broken test cases. In the case of larger test suites, with fewer changes, the cost of execution may become higher than repairing the test scripts. Another limitation of the existing techniques is that they are language dependent. For instance, S18 and S21 repairs QTP and Selenium IDE test scripts respectively. Both are capture-and-replay tools and share common characteristics (for example, capture the steps of actions on the web application user interfaces, which can later be replayed). Generic tools need to be developed to automatically provide test repairs, independent of the underlying testing framework for wider applicability. Interestingly, all 15 tools that we found were developed in java.

Table 4 Test case repair tools

| Name | Study | Type | Type of modification made to repair scripts | Test Case Execution | Domain | Available | Year |
|---|---|---|---|---|---|---|---|
| **REST** | S34 | GUI Test Scripts | Guide test personnel through changes in test scripts | Yes | GUI based applications | No | 2008 |
| **GUIAnalyzer** | S2 | GUI Test Scripts | Update GUI event sequences | No | GUI based applications | No | 2009 |
| **ReAssert** | S23 | JUnit | Replace literal values in tests, changing assertion methods, or replacing one assertion with other | Yes | Desktop applications | Yes | 2009 |
| **Test Care Assistant** | S19 | JUnit | Compilation errors lead certain changes in the method declaration | Yes | Desktop applications | Yes | 2011 |
| **WATER** | S21 | Selenium | Suggest repairs for assertion failures and element disposition | Yes | Web applications | No | 2011 |
| **TESTEVOL** | S5 | JUnit | Change method sequences and assertions values | Yes | Desktop applications | Yes | 2012 |
| **Maintenance Tool** | S8 | GUI Test Script | Take user feedback to repair test scripts | Yes | GUI based applications | No | 2012 |
| **FlowFixer** | S10 | GUI Test Scripts | Suggest replacement actions | Yes | GUI based applications | No | 2013 |
| **TestFix** | S4 | JUnit | Generate values to pass assert statements | Yes | Desktop applications | No | 2014 |
| **SITAR** | S18 | QTP | Use repairing transformations and human input | No | GUI based applications | No | 2016 |
| **WATERFALL** | S20 | Selenium | Suggest repairs for assertion failures and element disposition | Yes | Web applications | No | 2016 |
| **ATOM** | S9 | GUI Test Script | Update GUI event sequences | No | Mobile applications | No | 2017 |
| **VISTA** | S37 | Selenium | Suggest repairs for the broken test flow in the same page, widget shifted to neighboring page or removed. | Yes | Web applications | Yes | 2018 |
| **CHATEM** | S38 | GUI Test Script | Update GUI event sequences | No | Mobile applications | No | 2018 |
| **ITRACK** | S39 | JUnit | Repair broken method calls by using the replacing entities | No | Desktop applications | No | 2017 |

## 3.4. Test Frameworks (RQ 1.4)

This research question identifies the type of test scripts and testing frameworks which are mostly targeted by the repair approaches. Overall, 14 approaches repair test cases generated via JUnit framework, six repair approaches targeted Selenium scripts for repairing and two repair approaches targeted QTP scripts. For example, S3 presented an approach for repairing JUnit test cases and S8 for repairing Selenium test cases. Similarly, the approach presented in S14 repair test cases generated through QTP. There are 17 approaches categorized in ‘others’ that have not mentioned any specific target framework, nor could we infer the target framework from the paper. For example, S2 did not mention that the proposed approach repairs test cases for any particular target framework. Therefore, such approaches are categorized in ‘Others’. Figure 5 shows the type of test scripts repaired by the proposed approaches in the included studies. So far, JUnit is the most popular test framework targeted by most of the approaches. Figure 6 shows the distribution of platforms targeted by the included papers. Most of the published works focus on test breakage prevention and test repair of desktop applications. We found 7 papers that target web applications and 2 papers that explicitly cover mobile applications. Another 2 papers could not be placed in any of the categories due to their generic nature and lack of information that could be extracted. These are therefore mapped to others category. The two papers targeting test case repair for mobile applications target ROBOT test framework. Both the papers are from the same group of researchers. We did not find any works with other mobile application testing frameworks such as Appium, etc.

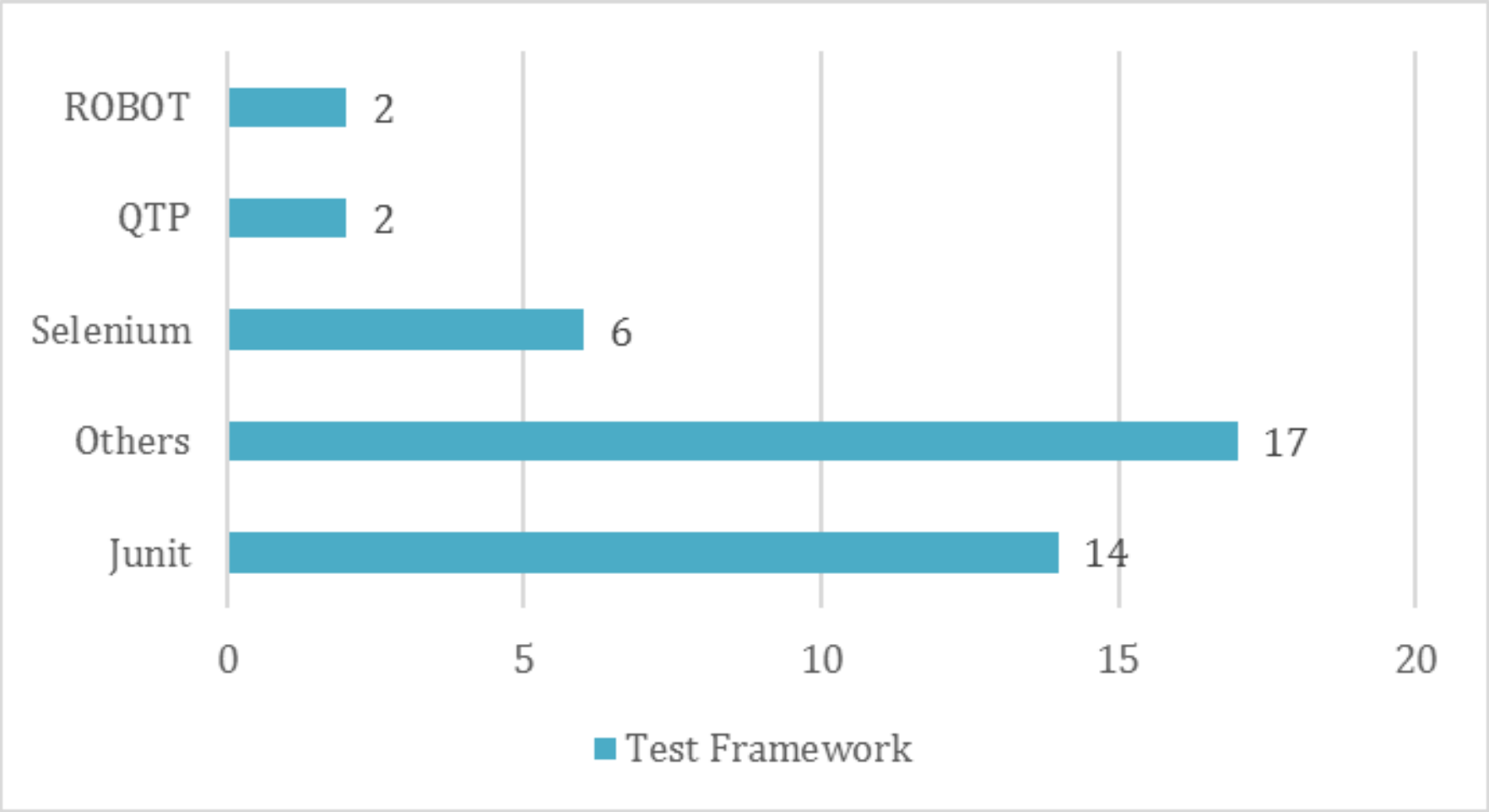

Figure 5 Type of test framework vs. number of papers

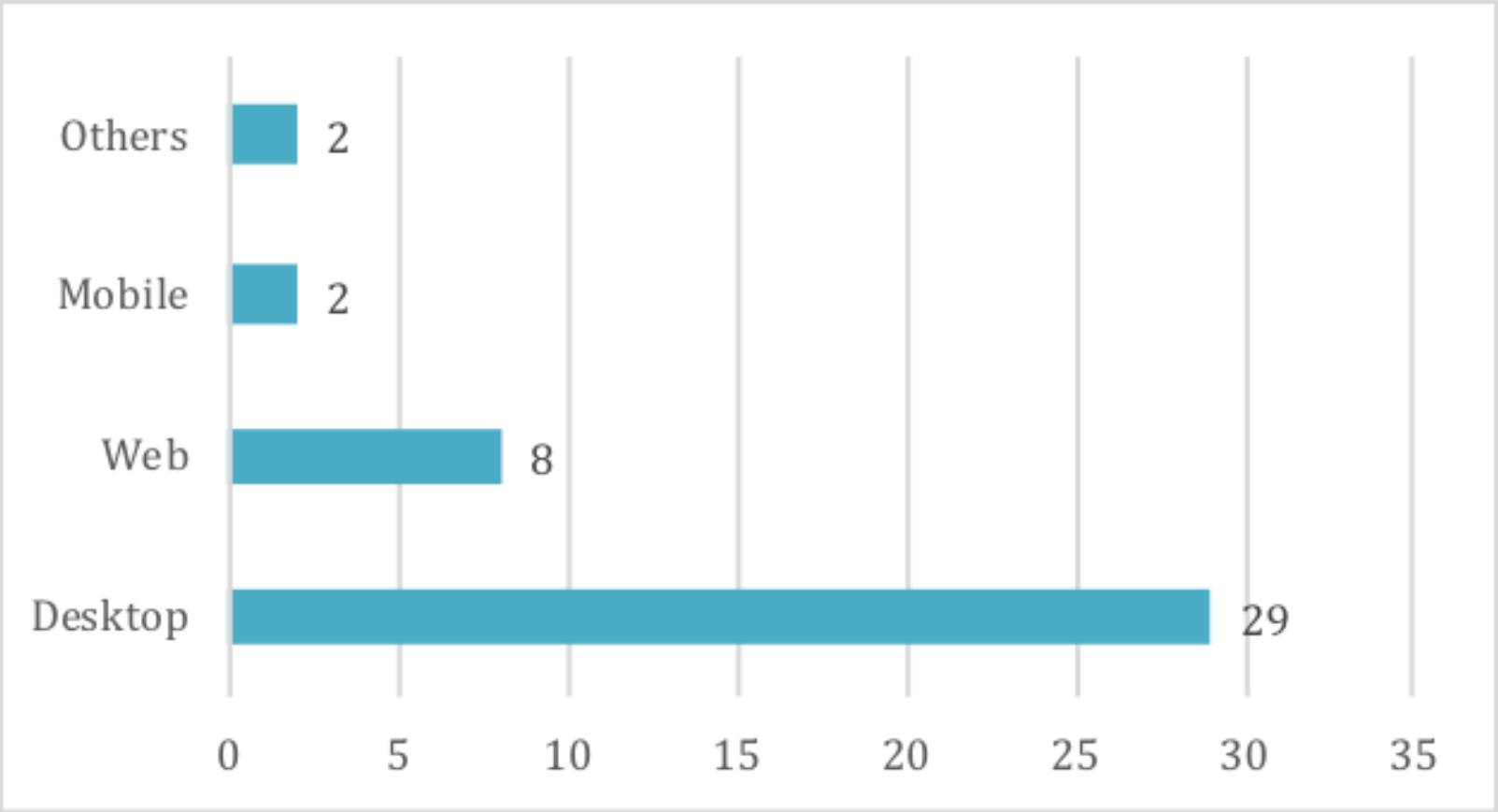

Figure 6 Type of test scripts vs. target platform

## 3.5. Automation Level (RQ 1.5)

Automation techniques are important for reducing test case repair efforts. We have assessed the level of automation of the existing techniques and have classified them as manual, semi-automatic and automatic, shown in Figure 7. By automated, we mean that the tool should automatically perform detection of test breakages and automatically generating potential test fixes. These fixes may be validated manually. There are twelve studies (30%) that provide automation of test case repair and were classified as automated approaches. For example, S21 proposes WATER tool to automatically suggest repairs for broken test scripts of web applications.

Techniques that contain manual steps, such as manual construction of the model in their approaches are classified as semi-automated, for example, S9 developed a tool, ATOM that requires manual construction of the event sequence model (ESM). This approach can be challenging at times as it requires knowledge about not only the changes, but also how the base version application. Nineteen studies (46%) provided semi-automated techniques. There were 10 studies (24%) that presented manual techniques for test case repair and were classified as manual. For example, S22 presented a technique that directs the testers in manually repairing broken test sequences for GUI. Such techniques require lots of human effort and time for repairing broken scripts. To summarize, numerous techniques have proposed automated techniques in the existing literature but mostly semi-automated techniques have been proposed that contain manual steps in their approaches.

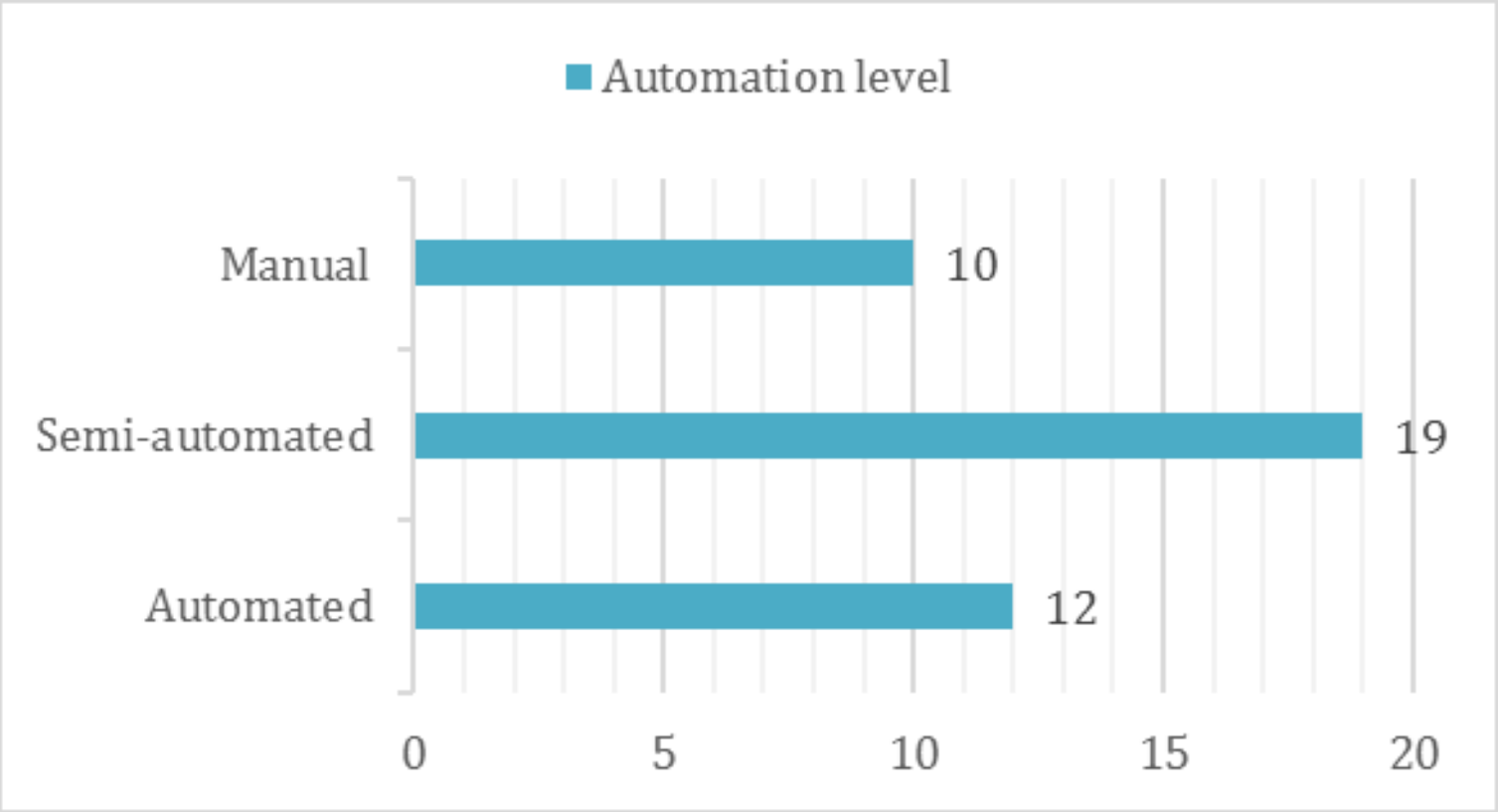

Figure 7 Type of automation vs. number of studies

## 3.6. Type of approaches for test repairs (RQ 1.6)

Our results indicate that most of the studies 36% (15) used model-based approaches in their test repair techniques, for example S22 has used a control flow graph to model the event sequence of the GUIs of the original version and the modified version to identify the changes and to check whether a test case is usable on modified GUI or not. About 34% (14 studies) have used symbolic and concolic execution approaches, for example, S23 has used dynamic symbolic execution to modify the values of assertions to make the test case pass. About 19% (8) studies used heuristic-based approaches, for example, S2 has provided some heuristics to solve the problem of maintaining GUI test cases. Five studies have used search-based approaches, for example, S4 has used the genetic algorithm for fixing broken JUnit tests.

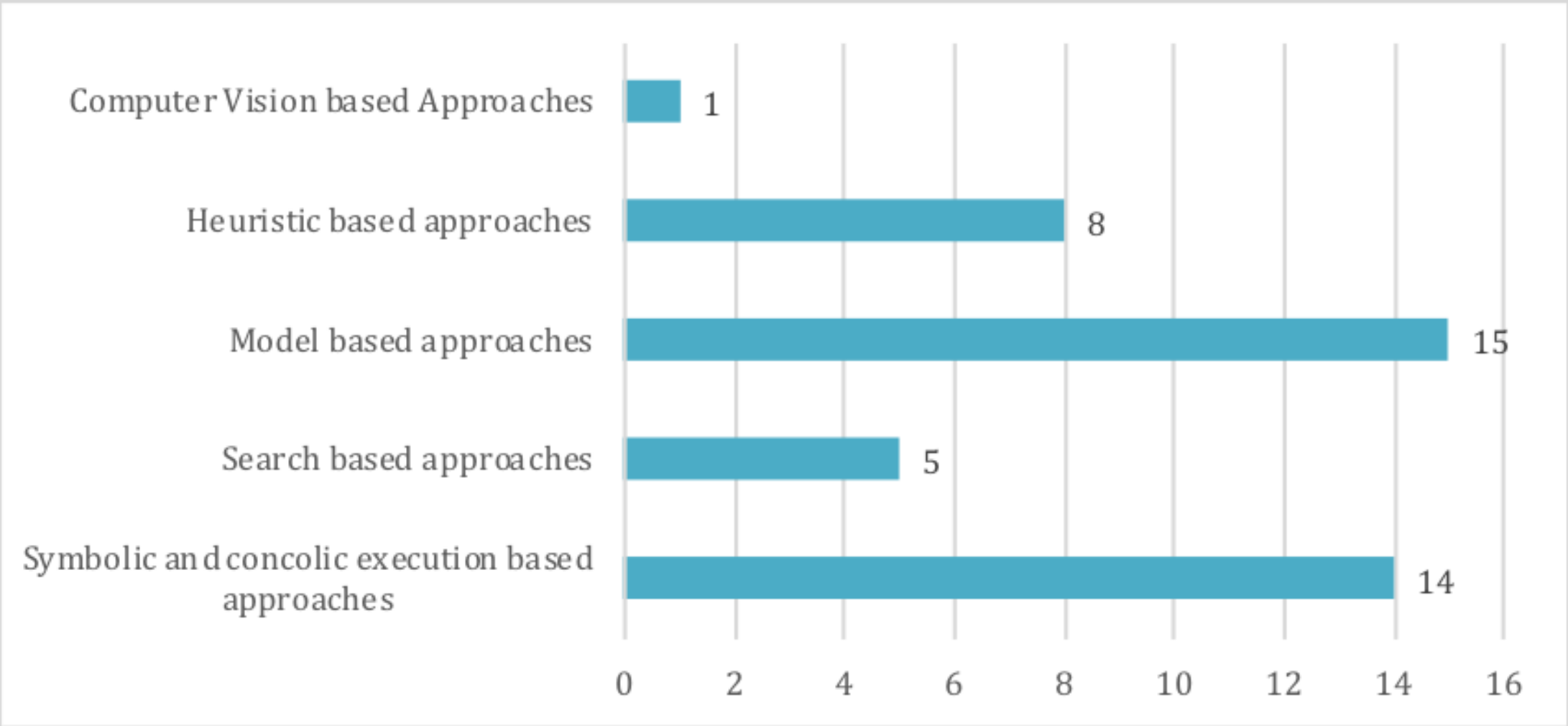

Figure 8 Type of approaches and percentages

Some studies use more than one approach, for example, S22 uses a model-based approach and propose heuristics to repair test cases. Recently a new paper (S37) is published that uses computer vision-based approach for repairing the GUI test script. Overall, model-based and symbolic execution-based approaches are the most popular approaches used by studies in their test repair techniques, as shown in

Figure *8*. Table 5 shows the summary of common weaknesses and strengths of the approaches used in the studies.

Table 5 Approaches strengths and weaknesses

| Approach | Strengths and Weaknesses |
|---|---|
| **Model-based approach** | + Ensure generalizability and provide tool-independent solutions [6].<br>- Requires expertise to design models [19]. |
| **Search-based approach** | + Provide the most optimized solution to the problem [20].<br>- Computationally expensive [21]. |
| **Heuristic-based approach** | + Can be used with any other repair techniques [22].<br>- May not provide accurate and generalize solution [23].<br>- It needs practitioner's experience and knowledge to apply heuristics efficiently [24]. |
| **Computer Vision-based approach** | + Visual locators might be the best choice when the visual appearance is more stable than the structure [25].<br>- Image processing algorithms are known to be quite computation-intensive and often reported as one of the weaknesses of visual testing [26]. |
| **Symbolic & Concolic based approach** | + Explore different feasible paths [5].<br>- These approaches are affected by path explosion problem [27]. |

## 3.7. Causes of test case breakages (RQ 2)

Software systems undergo several changes during their evolution. Unfortunately, such changes might affect the corresponding test cases. Some studies are available in the literature which has classified the causes of test breakages. For example, [28] provides a taxonomy of the causes of record and reply test breakages for evolving web application. Record and Replay tools record the interaction with the web browser while performing specific tasks. However, they are vulnerable to changes and will break during the test execution [29]. We extract the causes of test case breakages and collect them into a single taxonomy. Our taxonomy subsumes the taxonomy of causes of test case breakages for web applications presented in [28] and covers both desktop applications and mobile applications based on the data extracted from 41 included studies.

Such taxonomies help researchers to guide their test repair techniques for repairing maximum causes of breakages. It also helps to evaluate the maturity of approaches and clarifying key issues in the area. Without knowing the causes of broken test cases, it would not be possible to propose new approaches to repair them. Figure 9 shows the common causes of test breakages in all domains. The most common types of changes are the addition, deletion or modification of elements.

We summarize all identified causes of test breakages from the existing literature and re-classify them into coarse-grained classes on the basis of similarities among the causes, as shown in Table 6, Table 7 and Table 8. In order to integrate and classify the existing causes of test case breakages, two of the authors of this paper studied the presented causes in selected studies (where applicable) and labeled each cause of test breakage for creating the taxonomy independently. Subsequently, these labels were then refined through multiple review and group meetings of all authors for organizing them into hierarchies. For example, S8 is focused towards repairing the changes related to the method signature, class hierarchies and addition or deletion of overridden/overloaded methods. The targeted changes from each study were identified and grouped in some high-level classes with the consensus of all authors of the study.

Table 6 presents the type of code changes which can break their corresponding test cases. S6 is the only study which provides test repairs for almost all of the changes mentioned in the Table. S23 and S24 mostly deal with breakages related to method-level changes. We can conclude that “Method-Level Changes” (such as changes in the declaration of method parameters and return values, insertion and removal and type changes) are the prominent causes of test breakages for desktop applications. We did not find any work that focusses specifically code level changes for mobile and web applications. However, due to the nature of the approaches, it can be inferred that mobile and web applications will also share the same causes. Therefore, the approaches that fix and repair test breakage based on such changes should also be applicable to mobile and web applications.

Table 7 shows the causes of test breakages for testing web applications. Most of the techniques provide repairs for the broken HTML locators (such as id, name and XPath) and it also shows that web locators are a prominent cause of web test breakages. Changes related to pop-up boxes, page reloading and session expiry are neglected by the web test repair techniques. Table 8 shows the causes of test breakages due to GUI evolution of software systems. A number of test repair techniques provide fixes for the structural GUI evolution such as repositioning of GUI elements, enable or disable buttons, and other GUI layout changes. We found some instances of overlap between the causes of test breakages in this category between desktop, mobile and web applications. For example, the changes classified under event-related changes are common for all three platforms. Similarly, repositioning of graphical elements is also common to all three platforms. We did not find any works that address repairing of test cases broken as a result of session related changes (for example, user inactivity time increased or decreased), changes to Java scripts pop-ups, etc., despite being common in web applications.

As a consequence of classifying the reported causes in higher level classes, we hope to let researchers and practitioners infer whether a given technique may be applied on a SUT from a domain for which it was not originally intended. For example, S9 uses a model-based approach to repair broken GUI test scripts of mobile applications. This approach constructs an event sequence model (ESM) to abstract possible event sequences in a GUI and a delta ESM (DESM) to abstract the changes made to a GUI. By using delta DESM, it automatically updates test cases for the updated version. The use of modeling methodology makes it generalizable to be applied to other GUI-based desktop applications. VISTA (S37) uses computer vision-based approach to suggest repairs for broken capture and replay test scripts. As capture and replay test scripts share common characteristics like <locator, value, action>, this technique can also be used to repair other automated test scripts such as Selenium and QTP. Similarly, code level techniques such as those proposed by S4, S5, S6, S23 and S24 can also be applied mobile applications, web applications as well as desktop applications, even though the papers themselves do not provide any such application evidence.

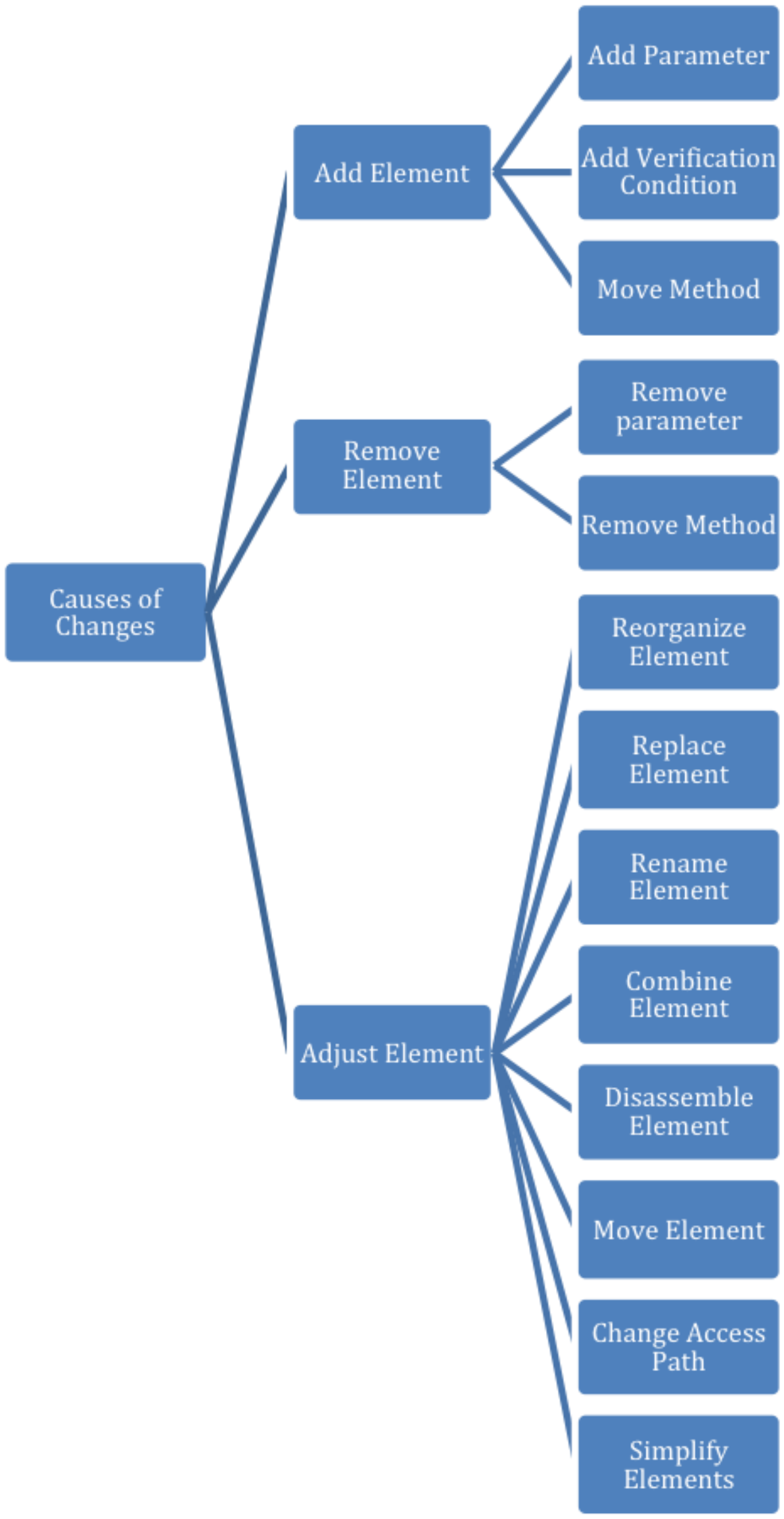

Figure 9: Common causes of test breakages

Table 6: Common causes of code level test breakages

| Level | Code-Level breakages | Platform | | |
|---|---|---|---|---|
| | | Desktop | Mobile | GUI |
| **Class-Level Changes** | **C1:** Add New Class<br>**C2:** Remove Class<br>**C3:** Change Class Type (e.g. static, normal, abstract etc.)<br>**C4:** Rename Class<br>**C5:** Combine Class<br>**C6:** Interface Implementation<br>**C7:** Extension of Class Hierarchy<br>**C8:** Update Class Hierarchy | S6 | Not Any | Not Any |
| **Method-Level Changes** | **C9:** Add New Method<br>**C10:** Method Parameter Added<br>**C11:** Method Parameter Deleted<br>**C12:** Add new condition<br>**C13:** Add overloaded method<br>**C14:** Add overridden method<br>**C15:** Assertion added<br>**C16:** Expected value modified<br>**C17:** Method Call deleted.<br>**C18:** Change Method Declaration<br>**C19:** Change Method Parameters type<br>**C20:** Change Method Return Type<br>**C21:** Change number of parameters<br>**C22:** Method Type Conversion (e.g. abstract, interface etc.)<br>**C23:** Change Access Specifier<br>**C24:** Merge Methods<br>**C25:** Move Methods | S4, S5, S6, S23, S24, S28, S39 | Not Any | Not Any |
| **Attribute-Level Changes** | **C26:** Add Attribute<br>**C27:** Delete Attribute<br>**C28:** Rename Attribute<br>**C29:** Move Attribute<br>**C30:** Modify Attribute<br>**C31:** Change Attribute Type (e.g. static, const) | S6, S23, S24 | Not Any | Not Any |

Table 7: Common causes of web test breakage (modified from [28] )

| Level | Description | Ref |
|---|---|---|
| Locator based breakages | **C1:** Addition/Deletion of new web elements.<br>**C2:** Rename element<br>**C3:** Adjust the position of the element. (E.g. change access path, replace the element and move an element from location to other).<br>**C4:** Modification of web element attribute (E.g. id, href, Alternative text, name, type, value, class and on Click).<br>**C5:** Addition/Deletion/Modification of an ancestor of an element in the DOM tree (e.g. div).<br>**C6:** Addition/Deletion of element – (Unable to find specified indexed element). | S8, S19, S20, S28, S33, S36, S21, S37, S41 |
| Value/Action related Changes | **C7:** Adding Verification condition (the e.g. value used by previous test case as an input is no longer accepted for updated version).<br>**C8:** Adding new web element in next version. (E.g. some fields were optional in version V but are mandatory in version V'.)<br>**C9:** Modify drop-down list.<br>**C10:** Delete option from drop-down list.<br>**C11:** Modify code (e.g. unable to match/compare actual value with expected). | S37, S20, S21 |
| JavaScript Popup boxes | **C12:** The absence of expected popup box.<br>**C13:** Presence of unexpected popup box. | Not targeted by any paper |
| Page Reloading | **C14:** Modify code (lack of time delays sufficient to allow its next version to succeed).<br>**C15:** User session timeout because of shorter time. | Not targeted by any paper |
| Session related Changes | **C16:** User inactivity time is increased in version V'.<br>**C17:** User inactivity time is decreased in version V'. | Not targeted by any paper |

Table 8: Common causes of GUI-related changes

| Level | Description | Platform | | |
|---|---|---|---|---|
| | | Desktop | Mobile | Web |
| Event-related changes | **C1:** Events cannot be dispatched once triggered.<br>**C2:** Actions that look similar but have different results.<br>**C3:** Different UI actions that may perform the same task.<br>**C4:** Presence or absence of confirmation modal dialog in an updated version.<br>**C5:** New action added, action deleted, action modified<br>**C6:** The execution time of specific action/service is different in an updated version. | S1, S10, S28 | S9 | S27 |
| Structural Changes | **C7:** Buttons become disabled due to some action.<br>**C8:** Deletion or relocation of elements<br>**C9:** Modify a button, add a button, and delete a button. | S1, S7, S9, S3, S18, | S9 | S27 |
| | **C10:** Identifier and text changes inside the visual hierarchy of activities.<br>**C11:** Layout and graphics change.<br>**C12:** Repositioning screen elements.<br>**C13:** Altering the selections in a drop-down list. | S16, S28, S22, S34, S29, S14, S25 | | |

## 3.8. System under Test (RQ 3.1)

As discussed in section 3, and shown in Table 9, we have extracted the following attributes for applications used in empirical evaluations.

a) Name of SUT
b) LoC size of SUT
c) Brief description of SUT
d) Frequency of the SUT used in studies
e) Type of the SUT, i.e. open source, experimental or commercial.
f) Language in which SUT is developed.
g) Domain

It can be noticed from the Table (highlighted in bold) that the largest case study used in the domain of mobile applications is Baidu Music having 5577 LoC. In web application Tikiwiki is the largest case study having 873000 LoC and in the experiment with the desktop applications, JFreeChart is used as a large case study having 217357 LoC.

Table 9 Characteristics of SUT

| S.no. | Name | Size (LOC) | Description | Frequency of usage as case study | Type | Language | Domain |
|---|---|---|---|---|---|---|---|
| **1** | PHP address book | 4000 | Web-based application for managing and organizing addresses and contacts. | 8 | Open source | PHP | Web |
| **2** | Collabtive | 68000 | Web-based software for managing geographically distributed teams to collaborate and work. | 8 | Open source | PHP | Web |
| **3** | PMD | 65279 | Static code analyser | 7 | Open source | Java | Desktop |
| **4** | **JFreeChart** | **217357** | **Chart generation library** | **5** | **Open source** | **Java** | **Desktop** |
| **5** | Xtream | 24655 | Download manager for increasing the download speed up to 500%. | 4 | Open source | Java | Desktop |
| **6** | MantisBT | 90000 | One of the most popular web-based bug tracking system | 4 | Open source | PHP | Web |
| **7** | Claroline | 277000 | A web-based collaborative e- | 4 | Open | PHP/MyS | Web |
| | | | learning application. | | source | QL | |
| **8** | Meeting room booking system | 9000 | Web-based application for reservation of rooms for meetings | 4 | Open source | PHP | Web |
| **9** | PHP password manager | 4000 | Web-based secured password manager | 5 | Open source | PHP | Web |
| **10** | Lucene | 1642 | Open source search engine | 3 | Open source | Java | Web |
| **11** | JodaTime | 63922 | Java date and time API | 3 | Open source | Java | Web |
| **12** | Joomla | 312978 | Content management system | 3 | Open source | PHP/MySQL | Web |
| **13** | CrosswordSage | 3220 | A tool to build professional crosswords with great word suggestion capabilities | 4 | Open source | Java | Desktop |
| **14** | FreeMind | 24665 | Mind mapping software | 3 | Open source | Java | Desktop |
| **15** | Common Lang | 5500 | Lang provides a host of helper utilities for the java. Lang API. | 2 | Open source | Java | Desktop |
| **16** | Common Math | 9550 | Java library for mathematics and statistics | 2 | Open source | Java | Desktop |
| **17** | Gson | 6500 | Java library used to convert java objects into JSON representation | 2 | Open source | Java | Desktop |
| **18** | Barbecue | 8842 | Java library for generation of barcode | 2 | Open source | Java | Desktop |
| **19** | Jedit | 5017 | Text editor | 2 | Open source | Java | Desktop |
| **20** | Gantt project | 3777 | Project management software | 2 | Open source | Java | Desktop |
| **21** | PHPFusion | 256899 | Light-weight content management system | 2 | Open source | PHP | Web |
| **22** | PHPAgenda | 43831 | A tool for managing appointments, holidays and to-do lists, etc., | 2 | Open source | PHP | Web |
| **23** | Dolibar | 42010 | Web-based Enterprise and CRM software | 2 | Open source | PHP | Web |
| **24** | TerpPaint | 13315 | Paint program with clipboard operations | 1 | Open source | Java | Desktop |
| **25** | TerpPresent | 44591 | An alternative to power point application | 1 | Open source | Java | Desktop |
| **26** | TerpWord | 22806 | An alternative to Microsoft word | 1 | Open source | Java | Desktop |
| **27** | TerpSpreadSheet | 6337 | A spreadsheet program with cells and tables | 1 | Open source | Java | Desktop |
| **28** | Ant | 93800 | Java library for driving processes describe in build files | 1 | Open source | Java | Desktop |
| **29** | Maven | 105100 | Software Project management tool based on project object model. | 1 | Open source | Java | Desktop |
| **30** | Strut | 110200 | Framework for java-based web applications | 1 | Open source | Java | Desktop |
| **31** | Spring Framework | 183100 | A framework for developing java applications | 1 | Open source | Java | Desktop |
| **32** | Handicapp | 3403 | A tool for listening and displaying the pronounced words of a speaker | 1 | Open source | Java | Mobile |
| **33** | Toile 2 Vert | 3389 | A tool for finding bike point, to recharge an electric bike | 1 | Open source | Java | Mobile |
| **34** | BiliBili | 1844 | Application for sharing video | 2 | Open source | Java | Mobile |
| **35** | Gnotes | 1489 | Simple notes application | 2 | Open source | Java | Mobile |
| **36** | Wannianli | 2397 | A simple calendar application | 2 | Open source | Java | Mobile |
| **37** | YoudaoNote | 3200 | Cloud based note tool | 2 | Open source | Java | Mobile |
| **38** | Wechat Phonebook | 3532 | Phone book application | 2 | Open source | Java | Mobile |
| **39** | ChangBa | 2800 | Karaoke application | 2 | Open source | Java | Mobile |
| **40** | **Baidu Music** | **5577** | **Music player** | **2** | **Open source** | **Java** | **Mobile** |
| **41** | 365 calender | 1207 | Calendar application | 2 | Open source | Java | Mobile |
| **42** | Ctrip | 4400 | Online travel agent | 2 | Open source | Java | Mobile |
| **43** | WizNote | 4936 | Cloud based IMS | 2 | Open source | Java | Mobile |
| **44** | TickTick | 1750 | To-do list application | 2 | Open source | Java | Mobile |
| **45** | JabRef | 38992 | Reference management system | 1 | Open source | Java | Desktop |
| **46** | JMSN | 11290 | Java Microsoft MSN clone | 1 | Open source | Java | Desktop |
| **47** | Twister | 492 | Application that allow users to write programs and download stock quotes | 1 | Open source | C# | Web |
| **48** | mRemote | 538 | Application for managing remote connections | 1 | Open source | C# | Web |
| **49** | University directory | 920 | Application that provides information about different university | 1 | Open source | C# | Web |
| **50** | Budget tracer | 343 | Software for tracking budget categories | 1 | Open source | C# | Web |
| **51** | Jmol | 2800 | Software for molecular modelling and chemical structures | 1 | Open source | Java | Desktop |
| **52** | AdblockIE | 2400 | Ad blocker for Internet Explorer | 1 | Open source | C# | Web |
| **53** | CSHgCmd | 2740 | C# interface to mercurial | 1 | Open source | C# | Web |
| **54** | Fudg-Csharp | 3800 | Binary message encoding | 1 | Open source | C# | Web |
| **55** | GCalExchangeSync | 7300 | Google calendars along with exchange server interoperability | 1 | Open source | C# | Web |
| **56** | Json.Net | 4350 | JSON serialization | 1 | Open source | C# | Web |
| **57** | MarkdounSharp | 2250 | Text to HTML convertor | 1 | Open source | C# | Web |
| **58** | NerdDinner | 3900 | A website for lunch plan | 1 | Open source | C# | Web |
| **59** | NGChart | 2800 | Wrapper for google charts API | 1 | Open source | C# | Web |
| **60** | Nhaml | 4900 | Template system for XHTML | 1 | Open source | C# | Web |
| **61** | ProjectPilot | 6200 | Source code statistics and metrics | 1 | Open source | C# | Web |
| **62** | SharpMap | 8800 | Geopatial mapping | 1 | Open source | C# | Web |
| **63** | FreeCol | 95404 | 4X video game | 1 | Open source | Java | Desktop |
| **64** | **TikiWiki** | **873000** | **Wiki-CMS-Groupware solution** | **1** | **Open source** | **PHP** | **Web** |
| **65** | OrangeHRM | 207000 | HR management system | 1 | Open source | PHP | Web |

Figure 10 shows the histogram of the 34 studies, which have conducted empirical evaluations and the number of subject applications they have used. S24 used the most number (17) of subject applications in its empirical evaluation. Furthermore, five studies (S5, S7, S12, S36, S41) uses six subject applications, four studies (S1, S15, S20, S35) uses seven subject applications, three studies (S3, S10, S40) uses five subject applications and three studies (S7, S12, S30) uses five subject applications in their empirical evaluations. Consequently, S33 uses eight subject applications, S6 uses nine subject applications, S9 uses 11 subject applications, S38 uses 16 subject applications and S24 uses 17 subject applications.

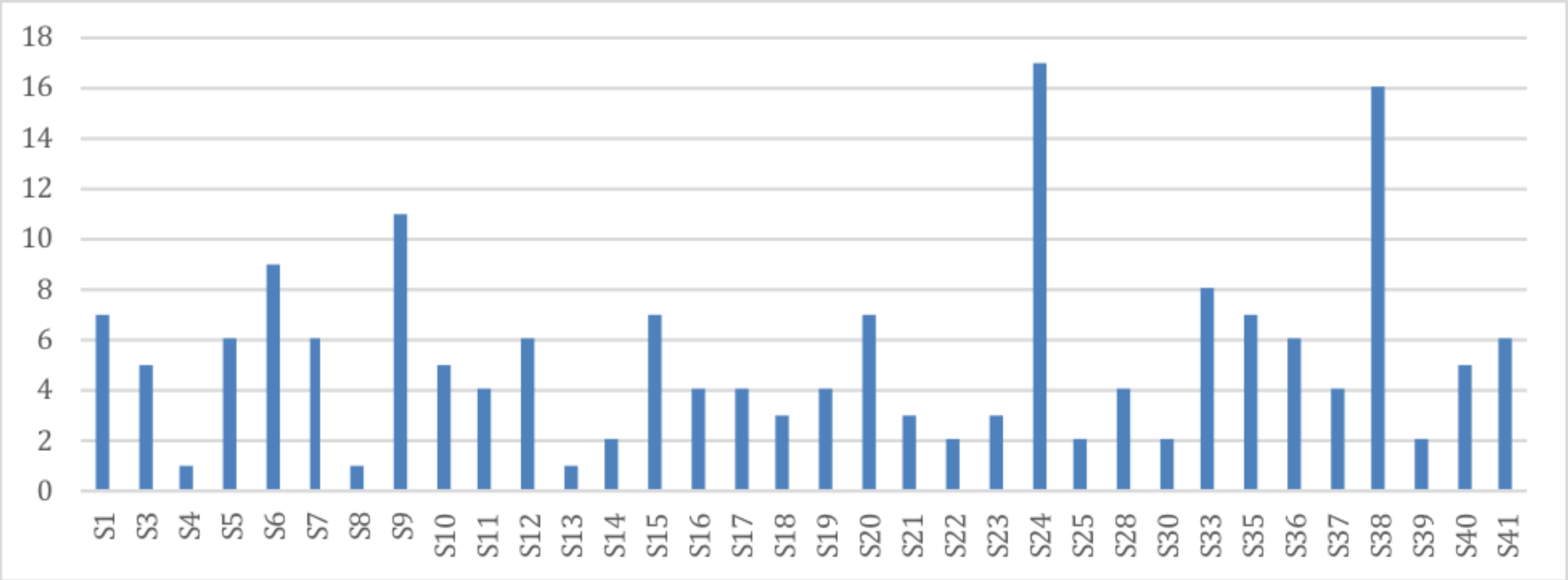

Figure 10 Number of case studies used by each study

Figure 11 shows the size (LOC) of SUT's used in each study. It is good to see that more than half (about 60%) of the studies used non-trivial SUT's (equal to or more than 10k) for evaluating their techniques. The study that uses largest SUT with 873000 LOC (named as TikiWiki) is S33, which was published in 2016.

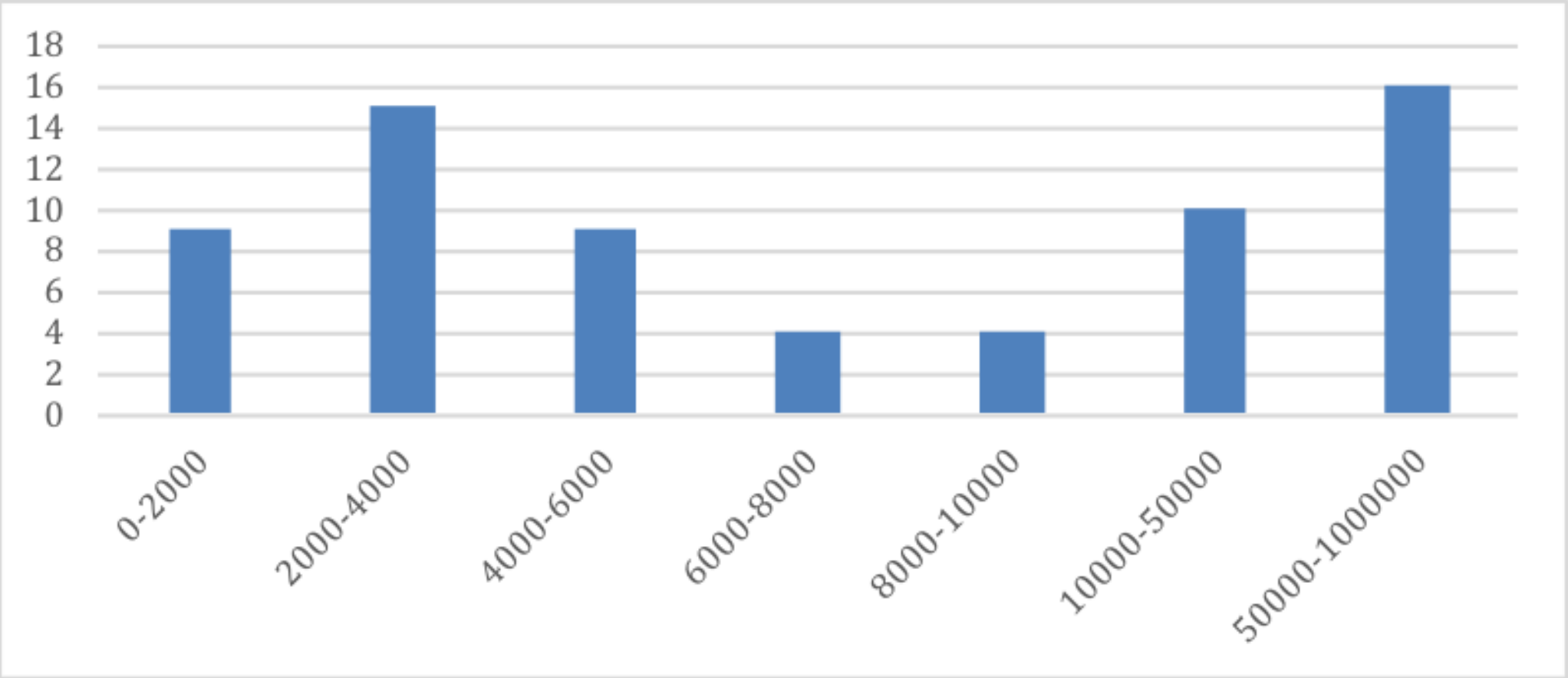

Figure 11 LOC of SUT's vs. Number of studies

We hypothesized that the size (LOC) may be increasing in new studies. To assess our hypothesis visually we have drawn a scatter plot, as shown in Figure 12, of years vs. size (LOC). Each dot in the Figure represents the LOC for each study w.r.t. year. In this context, we can argue that, in general, the size of SUT is increasing with time, i.e., newer papers are evaluating their approaches on multiple larger case studies which increases the confidence in their results.

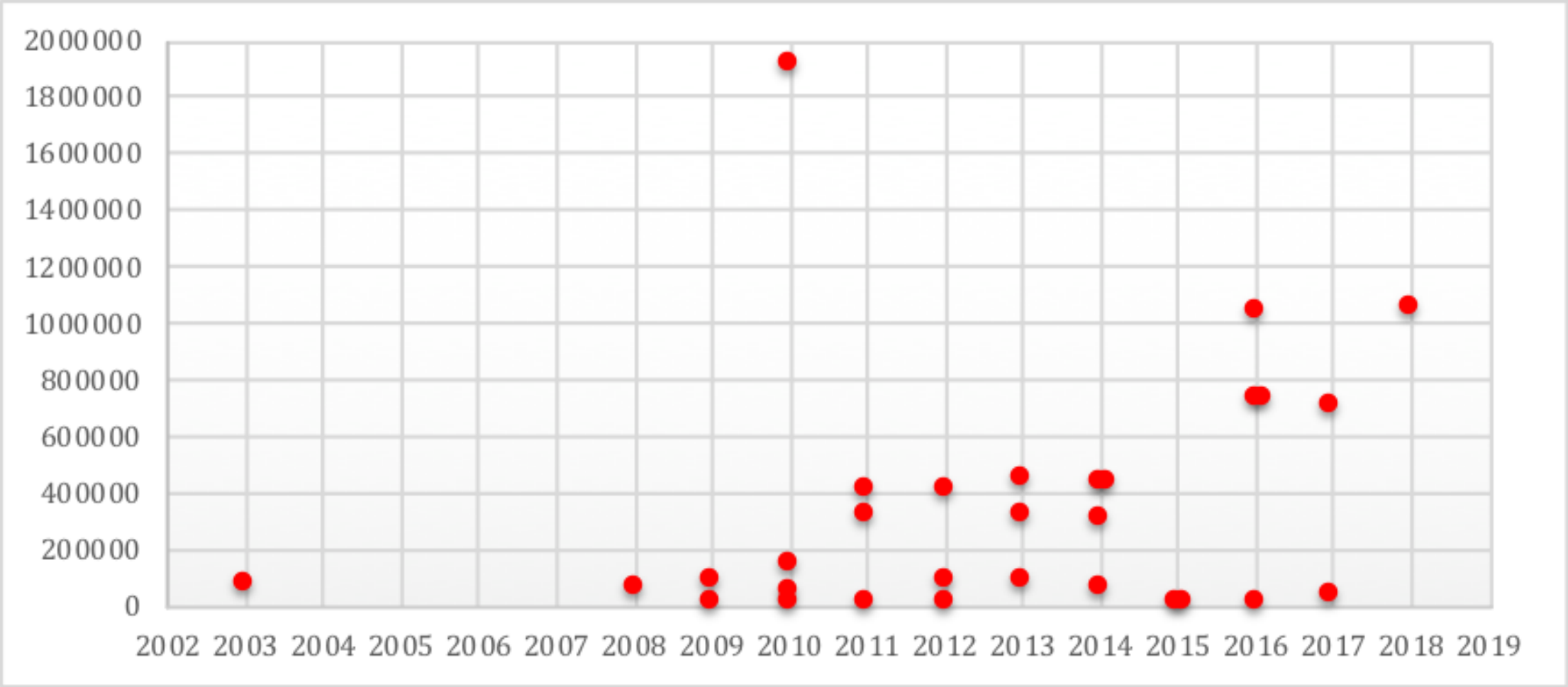

Figure 12 Years vs. LOC of SUT’s

Table 10 shows the list of most frequently used case studies and their respective number of downloads. The download statistics indicate (although imprecisely) that almost all frequently used case studies have a number of actual users. Therefore, these case studies are considered as industrial applications. Rigorous evaluation of test repair approaches on large case studies that have a number of actual users indicates the maturity of the area.

Table 10 Stats of frequently used case studies

| Frequently Used Case Studies | Usage Frequency | Last Update | Downloads | Total Commits |
|---|---|---|---|---|
| Php address book | 8 | Sep 11, 2016 | 153861 | 575 |
| Collabtive | 8 | Sep 19,2017 | 619729 | 153 |
| PMD | 7 | Jun 26,2018 | 11634080 | 272 |
| JFreeChart | 5 | Apr 13,2013 | 4404388 | 3646 |
| PHP password manager | 5 | July 9,2018 | 133235 | 435 |
| Xtream | 4 | July 16,2018 | 2370167 | 2590 |
| MantisBT | 4 | July 16,2018 | 2312232 | 11274 |
| Claroline | 4 | May 26,2018 | 356699 | 4965 |
| Meeting room booking system | 4 | Apr 15, 2012 | 439173 | 8 |
| Crossword Sage | 4 | July 9,2018 | 7565 | NA |
| Lucene | 3 | July 9,2018 | 8000 approx. | 30375 |
| joda-time | 3 | May 30, 2018 | 468558 | 2073 |
| Joomla | 3 | July 16, 2018 | 95000000 approx. | 30622 |
| FreeMind | 3 | Jun 28, 2015 | 23049691 | 66 |

## 3.9. Empirical evaluation metric (RQ 3.2)

Empirical evaluation of the proposed technique is essential for determining its applicability and suitability. To measure the effectiveness of test case repair technique, different studies have used multiple metrics. These metrics are used as criteria for assessing the effectiveness of the proposed test case repair technique. We have extracted these metrics from all 27 studies which have provided empirical evaluations. Table 11 shows the list of common metrics that studies have used/proposed for the purpose of measuring the effectiveness of test repair techniques. These metrics include test case length, test case execution time, code coverage (i.e. uniqueness of event sequences and branch coverage), screens and connection coverage (used mostly in the domain of mobile application testing) and false positives/negatives.

- **Test case length:** This metric is used to compute the length of test cases before and after the repair. In the context of GUI element coverage, longer test cases reduce the number of test cases necessary to achieve the test objectives [30]. This metric helps to evaluate the strength of approaches in maintaining the same number of steps in the test case. For example, a previous test case covers functionality by triggering four events/actions, but the repaired test case needs to trigger five events/actions for executing the complete test case. Hence, we can say that the repair test case has a maximum length than that of the previously broken test case.
- **Test case execution time:** This metric is used to estimate the execution time of the repaired test suite on the modified versions. A significant difference in resulted time when compared to the original test suite shows the strength or weakness of repaired approach. For instance, S28 measure the change in execution time of original test scripts of version 1.0 and compares it with the execution time of repaired test scripts of version 2.0.
- **Code coverage:** Studies used this metric to compare code coverage before and after the repair process. This metric measure if the test repair techniques improve or decrease the coverage of SUT.

- **Screens and connection coverage:** This metric is mostly used for mobile application where the intention is to test every screen (number of screens added/deleted) and connections (number of connections added/deleted/modified).
- **False positives and false negatives:** These metrics were mostly used in heuristics-based approaches. For instance, in the context of test case evolution, false negatives are the elements from the original GUI window identified as no longer existing in evolved GUI when in fact they actually do. False positives are the elements from the original GUI window which are identified to have been preserved (with little modifications) in evolved GUI but actually they are no longer present.

Table 11 Metrics used/proposed for the purpose of effectiveness measurement.

| Metrics | Paper Reference |
|---|---|
| **Test case length** | S1 |
| **Test case Execution time** | S1, S5 |
| **Code coverage** | S1, S5, S9, S18 |
| **Screens and connection coverage of mobile apps** | S9 |
| **False Positive, False Negative** | S2 |

### 3.10. Share of industry case studies (RQ 3.3)

Only a few studies have reported the use of 'real industrial case study' for evaluation of their proposed test case repair approaches. Majority of the studies with empirical evaluation have shown the applicability of their approaches by using open source applications. Some of these open source applications are close enough to the real industrial case study, for example, TikiWiki. However, there is a need for conducting empirical studies to assess the effectiveness of the proposed approaches on real industrial case studies.

## 4. Research findings

In this section, the findings of each research question are provided in the summarized form.

- **RQ 1.1 Type of research contribution**: Most of the research work is focused towards the contribution of new techniques or improving upon previous techniques on test repair. Fewer papers focussed on test breakage prevention. Of the published work, almost 55% of the studies have discussed test repair mechanisms whereas 31% of studies discussed test breakages and mechanisms to avoid test breakage. Only a few studies (12%) examined broken test case detection. Most of the published work can be classified as validation research and focuses on demonstrating the applicability of the proposed approach. We found 13 empirical evaluations that seek to establish evidence on the effectiveness of the approaches and identify which are the most effective approaches in a given context. These are discussed separately in the related works section. This information is critical for practitioners that are looking for a solution to the test suites maintenance problem for evolving applications. There is a lack of empirical evidence concerning which test repair tools are suitable for adaption by practitioners. We did not find any paper that reports the experience of using test repair in real industry projects. Similarly, there is little to no discussion on the cost of test repair, in particular comparison of automated test repair with semi-automated and manual approaches. Such a cost-effort analysis is an important factor for practitioners. There is a significant lack of controlled experiments and industry reports on whether test repair is feasible or not. Finally, where more than one tool is available for example for test repair of Selenium test cases, there is no evidence of which repair tools are most suitable or produce better results. Such results would be of great value to industry professionals.

- **RQ 1.2 Type of research method:** Majority of the studies were validated on open-source applications. The inclusion of large-scale open source case studies is a positive indication as such case studies are closer to industry applications. Additionally, empirical evaluations with open source subject application allow their replication. On the other hand, lack of reported results in real industry context reduces the confidence of practitioners on the maturity of proposed approaches. In particular, a number of approaches are presented as generic, without actual evidence of their application on specific case studies using a particular test framework. Researchers need to provide evidence on the application of their approaches and their feasibility using popular test frameworks. While we found a number of works that focus on JUnit and Selenium frameworks, other common testing frameworks such as APM, TestComplete, Espresso are not covered.

- **RQ 1.3 Test Case Repair Tools:** We discovered that most of the approaches target end-to-end (E2E) test scripts that operate at the GUI level and test the application as a whole from the point of view of the end user. However, from our observations, it emerges that most of the techniques have been proposed in the desktop domain, whereas the web and mobile domain are still understudied platforms. This is a positive indicator for practitioners involved in developing GUI based test cases. For researchers, we have identified a significant research gap in the domain of repairing web and mobile application test suites. Moreover, despite the claims of tools being publically available, at the time of our search, we only found a few tools that are still available for download. In particular, we did not find any tool available for download that can repair test cases of mobile applications. Such lack of tools is a significant hindrance in converting academic research into industry practices.

- **RQ1.4 Type of Test Framework:** We have identified the type of testing frameworks targeted by test repair approaches. Out of 11 tools reported in the literature, JUnit is the most frequent target framework for desktop applications. To our surprise, we only found 3 tools out of 15 that targets repairing of Selenium test cases, which is a popular open source testing framework for web applications. There were two reported tools that repair test cases of mobile applications by the same authors, repair test cases for the Robot test framework. No other testing framework for mobile applications is covered. This further strengthens the observation from RQ1.3 that there is a significant scope of industry application and experience papers. With a growing focus on automated testing of web applications, there is a critical need of tools that support test case repair to help in evolving the test suites.

- **RQ 1.5 Automation Level:** We identified that there is little or no empirical evidence on how these manual, semi-automated and automated approaches perform in terms of efficiency. Is it more efficient to repair the test cases or to simply throw them away and write new ones? Without empirical evidence on the effort required to repair test cases practitioners may be reluctant to adopt the proposed approaches. Furthermore, there is a significant scope of future research on automated validation of proposed test repairs. Any proposed repairs should not change the semantics of the test case. Currently, this validation is done manually.

- **RQ 1.6 Type of Approaches used in Test Repair Technique:** Model-based and symbolic execution-based approaches are the most popular approaches used by studies in their test repair techniques. Other approaches like search and heuristic based are less used in the area. In our opinion, an interesting future research direction could be to focus on how emerging data science technique can be applied in the area, particularly for applications that are being maintained for long time periods and consequently having a rich version history that may be used for mining.

- **RQ 2.1 Causes of Test Case Breakages:** We have provided a detailed taxonomy of changes/modifications in the applications which can break the existing test cases. We extracted test breakages repaired by each approach from papers and classified them into different classes and sub-classes. For a web application, the taxonomy presented in [28] is adopted. It is noticeable that researchers have proposed approaches for repairing specific test breakages in test suites. For practitioners, no tool is presented in the literature which repairs (close to) all identified changes/test breakages and that provides a generic solution. Most of the work tends to focus on desktop applications, there is a significant overlap in the causes of test case breakages between all three platforms. Consequently, it might be possible to apply some of the approaches proposed for repair desktop application test cases on test cases of mobile applications and vice-versa. This taxonomy can help researchers to improve existing approaches for test case repair and propose strategies according to the changes.

- **RQ 3.1 Characteristics of SUT:** We have found 27 studies which were evaluated on a wide range of web applications (SUTs) for their validations. This makes the tool or technique comparisons quite challenging in this field due to non-uniformity of the case studies. We have listed all subject applications used for empirical evaluation in the area. Most of the subject applications were developed in Java and their multiple versions and test cases are available. Most published empirical studies have used at least five or greater number of subject applications for the evaluation of their proposed approaches. We found a positive trend that more than half (about 60%) of the studies used large SUTs (equal to or more than 10k) for evaluating their techniques.

- **RQ 3.2 Evaluation Metrics for Empirical Studies:** We have identified different evaluation metrics from the empirical studies published in the area. This can help researchers and practitioners to effectively evaluate and compare different test case repair approaches based on the evaluation metrics.

- **RQ 3.3 Share of Industry Case Studies:** We have analysed the empirical studies in the area and found that no technique in the area is evaluated on an actual industrial case study. However, the most frequently used case studies are large scale open source studies that have a significant number of users (determined from the number of downloads). Therefore, the applications can be considered as good representative of industry applications. However, from a practitioner's perspective evaluation in real industry context is still an important aspect missing from the available literature. In particular, there is a lack of experience reports on challenges in applying these techniques on industry projects.

## 5. Related Work

To the best of our knowledge, there is currently no systematic review or literature survey in the area of test case evolution. However, numerous SLRs have been proposed in the area of software testing that we discussed in this section. Also, we discuss empirical studies in the area of test case repair.

### a. SLRs in Software Testing

There are a number of systematic reviews in different sub-areas of software testing [31]. For example, Dogan et al. [32] conducted an SLR on web application testing to identify, analyze and classify state-of-the-art techniques for testing of web applications. Kanewaka et al. [33] systematically gathered literature on the challenges and proposed solutions to testing of scientific software. Catal et al. [34] presented a systematic literature review on test case prioritization techniques using a genetic algorithm. The paper summarizes the existing techniques of the genetic algorithm for test case prioritization. Narciso et al. [35] conducted an SLR on the techniques of test case selection and state that random testing, genetic algorithm and greedy algorithm are the most commonly reported methods. Machado et al. [36] presented a systematic review on the strategies used in testing of software product lines. Rafi et al. [37] summarize the benefits and limitations of automated software testing by analyzing papers that presented techniques for test automation. Khan et al. [38] conducted an SLR on the reporting quality of model-based testing techniques.

### b. Empirical Studies in the area of Test Case Repair

We identified 13 empirical studies published in the area of test case repair, shown in Table 12. Here we provide a brief overview of existing secondary studies (e.g., empirical studies/taxonomy papers), focusing on different aspects of test case maintenance. For example, S42 presented an empirical analysis of Capture/Replay web testing and programmable web testing to evaluate their development time and test case maintenance effort. S43 presented a detailed taxonomy of causes of web test breakages. S44 conducted an empirical study to identify what costs are associated with automated GUI-based testing. S45 provides the fine-grained co-evolution patterns between production and test code. S46 has evaluated the feasibility of repairing broken test scripts automatically by studying maintenance operations on test scripts. S47 performed an empirical analysis to assess the robustness of visual and DOM-based web locators during code evolution. S48 presented an exploratory assessment to identify the causes of the fragility of UI automated tests for mobile applications. S49 studied the use of an optimal greedy algorithm to generate the robust XPath locators for web testing. S50 presented an extensive empirical study of the prevalence and maintenance of Selenium-based functional tests for web applications. S51 reported an experiment on an industrial case study, for investigating the potential benefits of adopting the page object pattern to improve the maintainability of Selenium WebDriver test cases. S52 has experimentally assessed the effectiveness of tool-based approach versus the manual approach for maintaining GUI directed test scripts. S53 presented the comparison of two test case generation algorithms (genetic and concolic) to examine the reuse of existing regression test cases by considering several factors (e.g. the order in which the code elements are targeted in the generation of test cases). S54 conducted an experiment to quantify the maintenance effort required to repair Selenium WebDriver test suites adopting different locators.

Table 12 Related Work: Empirical Studies

| ID | Author | Title | Year |
|---|---|---|---|
| S42 | Leotta et al. [39] | Capture-Replay vs. Programmable Web Testing: An Empirical Assessment during Test Case Evolution | 2013 |
| S43 | Hammoudi et al. [40] | Why Do Record/Replay Tests of Web Applications Break? | 2016 |
| S44 | Alegroth et al. [41] | Maintenance of automated test suites in industry: An empirical study on Visual GUI Testing | 2016 |
| S45 | Marsavina et al. [42] | Studying fine-grained co-evolution patterns of production and test code | 2014 |
| S46 | Christophe et al. [43] | Study on the Practices and Evolutions of Selenium Test Scripts | 2013 |
| S47 | Leotta et al. [25] | Visual vs. DOM-based web locators: An empirical study | 2014 |
| S48 | Coppola et al. [44] | Automated Mobile UI Test Fragility: An Exploratory Assessment Study on Android | 2016 |
| S49 | Leotta et al. [23] | Meta-Heuristic Generation of Robust XPath Locators for Web Testing | 2015 |
| S50 | Christophe et al. [45] | Prevalence and maintenance of automated functional tests for web applications | 2014 |
| S51 | Leotta et al. [46] | Improving test suites maintainability with the page object pattern: An industrial case study | 2013 |
| S52 | Grechanik et al. [47] | Experimental assessment of manual versus tool-based maintenance of GUI-directed test scripts | 2009 |
| S53 | Xu et al. [48] | Directed Test Suite Augmentation: An empirical investigation | 2009 |
| S54 | Leotta et al. [49] | Comparing the maintainability of selenium WebDriver test suites employing different locators: a case study | 2013 |

## 6. Threats to validity

In this section, threats to the validity of this SLR and the measures taken to minimize them are discussed.

*Internal threats validity:* One of the internal threats to this study is study selection. We have followed a systematic search process for searching papers and including them in our final selection. We have used different query strings to search in six major digital libraries for research papers and have used a rigorous inclusion and exclusion criteria for the final selection of our studies. Despite such a systematic process for the selection of studies, there are still chances of missing out some relevant study due to the way search strings are formed. There could also be studies published in languages other than English. We restricted our search only to manuscripts published in English.

*Researcher's bias* is another internal threat in the selection of primary studies. To reduce the threat each paper was reviewed by at least two authors of this study and all the conflicts in the selection of papers were discussed and resolved through multiple review and group meetings with all the authors of the study.

*External threats validity:* Generalizability in SLR can be interpreted as well the selected studies represent the area being studied. To ensure generalizability, we follow well-defined practices for conducting a systematic literature review and by including papers from all common databases and search engines. Snowballing was used on selected case studies to ensure that no studies are omitted that are relevant to the topic.

*Conclusion threat validity:* Conclusion validity of SLR deals with whether correct conclusions are drawn through systematic and repeatable treatments [16]. In order to confirm the reliability of the treatments, all the primary studies were reviewed carefully by at least two authors to reduce the bias in data extraction, which can lead to incorrect conclusions. Disagreements regarding the extracted data were resolved by consensus among the authors. The reported graphs and tables are directly generated from the extracted data in a spreadsheet to ensure its traceability with data. The systematic approach followed in this study ensures replicability and the results of any similar study will have no major deviations from our classification decisions. Additionally, we have made the extracted data available in an online Google spreadsheet (http://bit.do/eDsL6) for researchers to download and explore.

## 7. Conclusion

The goal of the study is to gather, analyze and classify the current state of the art in software test repair techniques. This review can help practitioners in many ways. It provides an overview of the state-of-the-art in the area and can be used as a catalog of existing test case repair techniques and tools. We have identified the published test case repair techniques, tools and their characteristics (metrics/attributes). Furthermore, it is found that researchers have focused on model-based and symbolic execution-based approaches to repair test cases. Most of the identified test repair techniques target test repair of GUI based applications. Despite the popularity of web applications and mobile applications, we found less research focus on test suites repair for mobile and web applications. Web and mobile applications represent a significant market share and more and more companies are moving towards automating their test suites. Techniques and tools that support test suites maintenance for web and mobile applications are therefore of significant interest to practitioners. Out of 15 proposed tools, available in the literature, we only found 4 tools that are publicly accessible. None of the tools proposed for repair test cases of mobile applications were available for download at the time of submission of this paper. Access to viable tools is an important consideration for the practitioners that are interested in evaluating and using a given approach. Without the tools being publicly available, it is difficult for a practitioner to evaluate its usefulness. Another important observation is on the nature of the case studies used for evaluating the proposed approaches. We found that empirical evaluations were done on open source case studies. We identified a positive trend of using large scale open source case studies for evaluation. Use of such large-scale case studies that have a number of active users increases the confidence in the results of the presented approach and is a loose indicator of growing maturity of research in the domain. Such studies would be a good approximation of industry case studies, we feel that researchers and practitioners would benefit significantly from experience reports and evaluation done in real industry settings. We found a significant lack of evidence on the comparison between the various tools and which tools and approaches are more suited for a given context. We found a number of manual, semi-automated and automated approaches that aim to repair test cases. However, we did not find evidence on the cost-effectiveness of such approaches. There is a need for controlled experiments and industry case studies to compare the effectiveness of the proposed test repair and breakage prevention approaches.

Table 13 List of selected studies

| ID | Author | Title | Year |
|---|---|---|---|
| S1 | Huang, Si [50] | A Framework for Automatically Repairing GUI Test Suites | 2010 |
| S2 | McMaster et al. [22] | An Extensible Heuristic-Based Framework for GUI Test Case Maintenance | 2009 |
| S3 | Gao et al. [51] | Analyzing Refactorings' Impact on Regression Test Cases | 2015 |
| S4 | Xu et al. [20] | Using Genetic Algorithms to Repair JUnit Test Cases | 2014 |
| S5 | Pinto et al. [52] | Understanding Myths and Realities of Test-suite Evolution | 2012 |
| S6 | Mirzaaghaei et al. [53] | Supporting Test Suite Evolution through Test Case Adaptation | 2012 |
| S7 | Huang et al. [54] | Repairing GUI test suites using a genetic algorithm | 2010 |
| S8 | Cunha and Maria Ana Casal [55] | Automatic maintenance of test scripts | 2011 |
| S9 | Li et al. [56] | ATOM: Automatic Maintenance of GUI Test Scripts for Evolving Mobile Applications | 2017 |
| S10 | Zhang et al. [57] | Automatically repairing broken workflows for evolving GUI applications | 2013 |
| S11 | Atif Memon [58] | Automatically repairing event sequence-based GUI test suites for regression testing | 2008 |
| S12 | Mirzaaghaei et al. [59] | Automatically repairing test cases for evolving method declarations | 2010 |
| S13 | Rapos et al. [60] | Examining the co-evolution relationship between Simulink Models and their test cases | 2016 |
| S14 | Priya et al. [61] | GUI Test Script Repair in Regression Testing | |
| S15 | Gove et al. [62] | Identifying infeasible GUI test cases using support vector machines and induced grammars | 2011 |
| S16 | Grechanik et al. [63] | Maintaining and evolving GUI-directed test scripts | 2009 |
| S17 | Yang et al. [64] | Specification-Based Test Repair Using a Lightweight Formal Method | 2012 |
| S18 | Gao et al. [6] | SITAR: GUI Test Script Repair | |
| S19 | Mirzaaghaei et al. [65] | TestCareAssistant: Automatic Repair of Test Case Compilation Errors | 2011 |
| S20 | Hammoudi et al. [66] | WATERFALL: an incremental approach for repairing record-replay tests of web applications | 2016 |
| S21 | Choudhary et al. [7] | WATER: Web Application TEst Repair | 2011 |
| S22 | Atif Memon and Mary Lou | Regression testing of GUIs | 2003 |
| | Soffa [3] | | |
| S23 | Daniel et al. [5] | ReAssert: Suggesting repairs for broken unit tests | 2009 |
| S24 | Daniel et al. [67] | On test repair using symbolic execution | 2010 |
| S25 | Atif Memon [68] | Using tasks to automate regression testing of GUIs | 2004 |
| S26 | Chen et al. [4] | When a GUI regression test failed, what should be blamed? | 2012 |
| S27 | Dhatchayani et al. [69] | Test case generation and reusing test cases for GUI designed with HTML | 2012 |
| S28 | Jiang et al. [70] | Assuring the model evolution of protocol software specifications by regression testing process improvement | 2011 |
| S29 | Daniel et al. [71] | Automated GUI Refactoring and Test Script Repair | 2011 |
| S30 | Hao et al. [72] | Is this a bug or an obsolete test? | 2013 |
| S31 | Mayan et al. [73] | Novel Approach to Reuse Unused Test Cases in a GUI Based Application | 2015 |
| S32 | Evans et al. [74] | Differential testing: A new approach to change detection | 2007 |
| S33 | Leotta et al. [75] | ROBULA +: an algorithm for generating robust XPath locators for web testing | 2016 |
| S34 | Xie et al. [76] | REST: A Tool for Reducing Effort in Script-based Testing | 2008 |
| S35 | Tan et al. [77] | relifix: Automated repair of software regressions | 2015 |
| S36 | Leotta et al. [25] | Reducing Web Test Cases Aging by means of Robust XPath Locators | 2014 |
| S37 | Stocco et al. [78] | Visual Web Test Repair | 2018 |
| S38 | Chang et al. [79] | Change-Based Test Script Maintenance for Android Apps | 2018 |
| S39 | Nguyen et al. [80] | Interaction-Based Tracking of Program Entities for Test Case Evolution | 2017 |
| S40 | Leotta et al. [81] | Using multi-locators to increase the robustness of web test cases | 2015 |
| S41 | Yandrapally et al. [82] | Robust test automation using contextual clues | 2014 |

**References:**

1. M. Mirzaaghaei, "Automatic Test Suite Evolution". 2012.
2. M. Leotta, D. Clerissi, F. Ricca, and P. Tonella. , "Chapter five-approaches and tools for automated end-to-end web testing". Advances in Computers 101 (2016): 193-237.
3. A.M. Memon and M.L. Soffa, "Regression testing of GUIs". Proceedings of the 9th European software engineering conference held jointly with 10th ACM SIGSOFT international symposium on Foundations of software engineering - ESEC/FSE '03, 2003: p. 118.
4. J. Chen, M. Lin, K. Yu, B. Shao, "When a GUI regression test failed, what should be blamed?". 5th International Conference on Software Testing, Verification and Validation, ICST 2012. 2012. p. 467-470.
5. B. Daniel, V. Jagannath, D. Dig, "ReAssert: Suggesting repairs for broken unit tests". International Conference on Automated Software Engineering, 2009: p. 433-444.
6. Z. Gao, Z. Chen, Y. Zou, "SITAR: GUI Test Script Repair". IEEE Transactions on Software Engineering, 2016. 42: p. 170-186.
7. S.R. Choudhary, D. Zhao, H. Versee, A. Orso, "WATER : Web Application TEst Repair". First International Workshop on EndtoEnd Test Script Engineering, 2011: p. 24-29.

8. B. Kitchenham, O. Pearl Brereton, D. Budgen, M. Turner, J. Bailey, and S. Linkman, "Systematic literature reviews in software engineering–a systematic literature review". Information and software technology 2009: p. no. 1 (2009): 7-15.
9. B. Kitchenham, S.L. Pfleeger, L.M. Pickard, "Preliminary guidelines for empirical research in software engineering". Software Engineering,. IEEE Transactions on, 2002. 28(8): p. 721-734., 2002.
10. R.V. Solingen, V. Basili, G. Caldiera, and H. D. Rombach. "Goal question metric (gqm) approach". Encyclopedia of software engineering (2002).
12. K. Petersen, R.Feldt., S. Mujtaba, M. Mattsson, "Systematic mapping studies in software engineering".International Conference on Evaluation and Assessment in Software Engineering (EASE), 2008, pp. 71–80. , 2008.
13. Muhammad Uzair Khan, Salman Sherin, Muhammad Zohaib Iqbal, and Rubab Zahid. "Landscaping systematic mapping studies in software engineering: A tertiary study." Journal of Systems and Software 149 (2019): 396-436.
14. H. Munir, K. Wnuk, and P. Runeson, "Open innovation in software engineering: a systematic mapping study". Empirical Software Engineering, 2016. **21**(2): p. 684-723.
15. K. Petersen, S. Vakkalanka, and L. Kuzniarz, "Guidelines for conducting systematic mapping studies in software engineering: An update". Information and Software Technology, 2015. 64: p. 1-18.
16. V. Garousi and M.V. Mäntylä, "A systematic literature review of literature reviews in software testing". Information and Software Technology, 2016. 80: p. 195-216.
17. C. Wohlin, "Guidelines for snowballing in systematic literature studies and a replication in software engineering". International conference on evaluation and assessment in software engineering. 2014. ACM.
18. L. Chen, M. A. Babar, and H. Zhang, "Towards an evidence-based understanding of electronic data sources". 2010.
19. J. Hutchinson, J. Whittle, M. Rouncefield, S. Kristoffersen, "Empirical assessment of MDE in industry". 33rd international conference on software engineering. 2011. ACM.
20. Y. Xu, B. Huang, G. Wu, M. Yuan, "Using genetic algorithms to repair JUnit test cases". Proceedings - Asia-Pacific Software Engineering Conference, APSEC, 2014. **1**: p. 287-294.
21. S. Ali, L.C. Briand, H. Hemmati, R.K.P. Walawege, "A systematic review of the application and empirical investigation of search-based test case generation". IEEE Transactions on Software Engineering, 2010. **36**(6): p. 742-762.
22. S. McMaster and A.M. Memon, "An extensible heuristic-based framework for gui test case maintenance." In Software Testing, Verification and Validation Workshops, 2009. ICSTW'09. International Conference on, pp. 251-254. IEEE, 2009., 2009.
23. M. Leotta, A. Stocco, F. Ricca, P. Tonella, "Meta-heuristic generation of robust XPath locators for web testing". Search-Based Software Testing (SBST), 2015 IEEE/ACM 8th International Workshop on. 2015. IEEE.
24. D. Ferguson, M. Likhachev, and A. Stentz, "A guide to heuristic-based path planning". International workshop on planning under uncertainty for autonomous systems, international conference on automated planning and scheduling (ICAPS). 2005.
25. M. Leotta, D. Clerissi, F. Ricca, P. Tonella, "Visual vs. DOM-based web locators: An empirical study". International Conference on Web Engineering. 2014. Springer.

26. E. Börjesson and R. Feldt. "Automated system testing using visual gui testing tools: A comparative study in industry". 2012 IEEE Fifth International Conference on Software Testing, Verification and Validation. 2012. IEEE.
27. C. Cadarand K. Sen, "Symbolic execution for software testing: three decades later". Communications of the ACM, 2013**.** 56(2): p. 82-90.
28. M. Hammoudi, G. Rothermel and P. Tonella, "Why do Record/Replay Tests of Web Applications Break?" 2016 IEEE International Conference on Software Testing, Verification and Validation (ICST), Chicago, IL, 2016, pp. 180-190. doi: 10.1109/ICST.2016.16
29. M. Leotta, D. Clerissi, F. Ricca, P. Tonella, "Approaches and tools for automated end-to-end web testing". Advances in Computers. 2016, Elsevier. p. 193-237.
30. S. Carino and J.H. Andrews. "Evaluating the effect of test case length on GUI test suite performance". 10th International Workshop on Automation of Software Test. 2015. IEEE Press.
31. K. Petersen, "Systematic Mapping Studies in Software Engineering". Evaluation and Assessement in Software Engineering. 2008.
32. S. Doğan, A.B. Can, and V. Garousi, "Web application testing: A systematic literature review". Journal of Systems and Software, 2014. 91: p. 174-201.
33. U. Kanewala and J.M. Bieman, "Testing scientific software: A systematic literature review". Information and software technology, 2014. 56(10): p. 1219-1232.
34. C. Catal, "On the application of genetic algorithms for test case prioritization: a systematic literature review". 2nd international workshop on evidential assessment of software technologies. 2012. ACM.
35. E.N. Narciso, M.E. Delamaro, and F.D.L.D.S. Nunes, "Test case selection: A systematic literature review". International Journal of Software Engineering and Knowledge Engineering, 2014. 24(04): p. 653-676.
36. I. do Carmo Machado, J.D. McGregor, Y.C. Cavalcanti, E.S. de Almeida, "On strategies for testing software product lines: A systematic literature review". Information and Software Technology, 2014. 56(10): p. 1183-1199.
37. D.M. Rafi, K.R.K. Moses, K. Petersen, M. Mantyla, "Benefits and limitations of automated software testing: Systematic literature review and practitioner survey". in Proceedings of the 7th International Workshop on Automation of Software Test. 2012. IEEE Press.
38. M. Uzair khan, S. Iftikhar, M.Z. Iqbal, S. Sherin "Empirical studies omit reporting necessary details: A systematic literature review of reporting quality in model based testing". Computer Standards & Interfaces, 2018. **55**: p. 156-170.
39. M.Leotta, D. Clerissi, F. Ricca, P. Tonella, "Capture-replay vs. programmable web testing: An empirical assessment during test case evolution". in Reverse Engineering (WCRE), 2013 20th Working Conference on. 2013. IEEE.
40. M. Hammoudi, G. Rothermel and P. Tonella. "Why do record/replay tests of web applications break?" in Software Testing, Verification and Validation (ICST), 2016 IEEE International Conference on. 2016. IEEE.
41. E. Alégroth, R. Feldt and P. Kolström, "Maintenance of automated test suites in industry: An empirical study on Visual GUI Testing". Information and Software Technology, 2016. **73**: p. 66-80.

42. C. Marsavina, D. Romano, and A. Zaidman. "Studying fine-grained co-evolution patterns of production and test code". Source Code Analysis and Manipulation (SCAM), 2014 IEEE 14th International Working Conference on. 2014. IEEE.
43. L. Christophe, C. De Roover, and W. De Meuter, "Study on the Practices and Evolutions of Selenium Test Scripts". BENEVOL 2013, 2013: pp. 13.
44. R. Coppola, E. Raffero, and M. Torchiano. "Automated mobile UI test fragility: an exploratory assessment study on Android". in Proceedings of the 2nd International Workshop on User Interface Test Automation. 2016. ACM.
45. L. Christophe, R. Stevens, C. De Roover, W.D. Meuter, "Prevalence and maintenance of automated functional tests for web applications". Software Maintenance and Evolution (ICSME), 2014 IEEE International Conference on. 2014. IEEE.
46. M. Leotta, D. Clerissi, F. Ricca, "Improving test suites maintainability with the page object pattern: An industrial case study". in Software Testing, Verification and Validation Workshops (ICSTW), 2013 IEEE Sixth International Conference on. 2013. IEEE.
47. M. Grechanik, Q. Xie, and C. Fu, "Experimental assessment of manual versus tool-based maintenance of gui-directed test scripts". International Conference in Software Maintenance (ICSM) 2009. IEEE International Conference on. 2009.
48. Z. Xu, Y. Kim, M. Kim, M.B. Cohen and G. Rothermel, "Directed test suite augmentation: An empirical investigation". Software Testing Verification and Reliability, 2009, 25 (2): pp. 77-114.
49. M.Leotta, D. Clerissi, F. Ricca, C. Spadaro, "Comparing the maintainability of selenium webdriver test suites employing different locators: A case study". in Proceedings of the 2013 international workshop on joining academia and industry contributions to testing automation. 2013. ACM.
50. S. Huang, "A Framework for Automatically Repairing GUI Test Suites". (2010).
51. Y. Gao, H. Liu, X. Fan, Z. Niu, and B. Nyirongo, "Analyzing Refactorings' Impact on Regression Test Cases.". In Computer Software and Applications Conference (COMPSAC), 2015 IEEE 39th Annual, , IEEE, 2015. vol. 2, pp. 222-231.
52. L.S. Pinto, S. Sinha, and A. Orso, "Understanding Myths and Realities of Test-suite Evolution". Proceedings of the ACM SIGSOFT 20th International Symposium on the Foundations of Software Engineering, 2012. **1**: pp. 33:1-33:11.
53. M. Mirzaaghaei, F. Pastore, and M. Pezz, "Supporting Test Suite Evolution through Test Case Adaptation". International Conference on Software Testing, Verification and Validation, 2012.
54. S. Huang, M.B. Cohen, and A.M. Memon, "Repairing GUI test suites using a genetic algorithm". ICST 2010 - 3rd International Conference on Software Testing, Verification and Validation, 2010: p. 245-254.
55. M.A.C. Cunha, "Automatic maintenance of test scripts". 2012.
56. X. Li, N. Chang, Y. Wang, H. Huang, Y. Pei, L. Wang, and Xuandong Li. , "ATOM: Automatic Maintenance of GUI Test Scripts for Evolving Mobile Applications". In Software Testing, Verification and Validation (ICST), 2017 IEEE International Conference on, IEEE, 2017.: p. pp. 161-171.
57. S. Zhang, H. Lü, and M.D. Ernst, "Automatically repairing broken workflows for evolving GUI applications". Proceedings of the 2013 International Symposium on Software Testing and Analysis - ISSTA 2013, 2013: p. 45.

58. A.M. Memon, "Automatically repairing event sequence-based GUI test suites for regression testing". ACM Transactions on Software Engineering and Methodology, 2008. **18**: p. 1-36.
59. M. Mirzaaghaei, F. Pastore, and M. Pezze. "Automatically repairing test cases for evolving method declarations". in Software Maintenance (ICSM), 2010 IEEE International Conference on. 2010. IEEE.
60. E.J. Rapos and J.R. Cordy, "Examining the co-evolution relationship between Simulink Models and their test cases". Proceedings - 8th International Workshop on Modeling in Software Engineering, MiSE 2016, 2016: p. 34-40.
61. G. Priya and B. Rao, "GUI Test Script Repair in Regression Testing". ermt.net.
62. R. Gove and J. Faytong, "Identifying infeasible GUI test cases using support vector machines and induced grammars". Proceedings - 4th IEEE International Conference on Software Testing, Verification, and Validation Workshops, ICSTW 2011, 2011: p. 202-211.
63. M. Grechanik, Q. Xie, and C. Fu, "Maintaining and evolving GUI-directed test scripts". Proceedings - International Conference on Software Engineering, 2009: p. 408-418.
64. G. Yang, S. Khurshid, and M. Kim, "Specification-Based Test Repair Using a Lightweight Formal Method". 2012: p. 455-470.
65. M. Mirzaaghaei and F. Pastore, "TestCareAssistant : Automatic Repair of Test Case Compilation Errors".
66. M. Hammoudi, G. Rothermel, and A. Stocco, "WATERFALL: an incremental approach for repairing record-replay tests of web applications". Proceedings of the 2016 24th ACM SIGSOFT International Symposium on Foundations of Software Engineering - FSE 2016, 2016: p. 751-762.
67. B. Daniel, T. Gvero, and D. Marinov, "On test repair using symbolic execution". Proceedings of the 19th international symposium on Software testing and analysis - ISSTA '10, 2010: p. 207.
68. A.M. Memon, "Using tasks to automate regression testing of GUIs". Conferernce on Artificial intelligence and Applications (AIA), 2004.
69. M. Datchayani, A.X.A. Rayan, Y.palanichamy, and B. Zacharias, "Test case generation and reusing test cases for GUI designed with HTML". Journal of Software, 2012. **7**: pp. 2269-2277.
70. B. Jiang, T.H. Tse, W. Grieskamp, N. Kicillof, Y. Cao, X. Li, W.K. Chan, "Assuring the model evolution of protocol software specifications by regression testing process improvement". Software: Practice and Experience, 2011. 41(10): pp. 1073-1103.
71. B. Daniel, Q. Luo, M. Mirzaaghaei, D. Dig, D. Marinov, and M. Pezze, "Automated GUI refactoring and test script repair". in Proceedings of the First International Workshop on End-to-End Test Script Engineering. 2011. ACM.
72. D. Hao, T. Lan, H. Zhang, C. Guo and L. Zhang, "Is this a bug or an obsolete test?" Lecture Notes in Computer Science (including subseries Lecture Notes in Artificial Intelligence and Lecture Notes in Bioinformatics), 2013. 7920 LNCS: pp. 602-628.
73. J, A.Mayan and K.L. Priya, "Novel Approach to Reuse Unused Test Cases in a GUI Based Application". International Conference on Circuits, Power and Computing Technologies. 2015.
74. R.B. Evansand A. Savoia, "Differential testing: A new approach to change detection". Aids, 2007: p. 549-552.

75. M. Leotta, A. Stocco, F. Ricca, P. Tonella, "ROBULA + : an algorithm for generating robust XPath locators for web testing". 2016: p. 177-204.
76. Q. Xie, M. Grechanik, and C. Fu, "REST : A Tool for Reducing Effort in Script-based Testing". International Conference on Software Maintenance, 2008.
77. S.H. Tan and A. Roychoudhury, "Relifix: Automated repair of software regressions". Proceedings - International Conference on Software Engineering, 2015. **1**: p. 471-482.
78. A. Stocco, R. Yandrapally, and A. Mesbah. "Visual Web Test Repair". Proceedings of the 26th ACM Joint European Software Engineering Conference and Symposium on the Foundations of Software Engineering (ESEC/FSE 2018). ACM. 2018.
79. N. Chang, L. Wang, Y. Pei, S.K. Mondal, and X. Li, "Change-Based Test Script Maintenance for Android Apps". 2018 IEEE International Conference on Software Quality, Reliability and Security (QRS). 2018.
80. H.A. Nguyen, T.T. Nguyen, T.N. Nguyen, H.V. Nguyen, "Interaction-Based Tracking of Program Entities for Test Case Evolution". Software Maintenance and Evolution (ICSME), 2017 IEEE International Conference on. 2017. IEEE.
81. M. Leotta, A. Stocco, F. Ricca, P. Tonella, "Using multi-locators to increase the robustness of web test cases". Software Testing, Verification and Validation (ICST), 2015 IEEE 8th International Conference on. 2015. IEEE.
82. R. Yandrapally, S. Thummalapenta, S. Sinha, S. Chandra, "Robust test automation using contextual clues". in Proceedings of the 2014 International Symposium on Software Testing and Analysis. 2014. ACM.